\documentclass[aps,preprint,noshowpacs,nofootinbib,letterpaper,eqsecnum,preprintnumbers]{revtex4}
\usepackage{amssymb,amsmath,graphicx}
\usepackage{hyperref}
\usepackage{color}

\newcommand{\non}{\nonumber \\}
\newcommand{\noi}{\noindent}

\newcommand{\alp}{\alpha}     
     
\newcommand{\eps}{\epsilon}

\newcommand{\kap}{\kappa}     
   \newcommand{\sig}{\sigma}
 
   \newcommand{\ome}{\omega}

\newcommand{\Gam}{\Gamma}     
     
\newcommand{\Sig}{\Sigma}

\newcommand{\cO}{{\cal O}}

\newcommand{\RR}{\mathbb{R}}

\newcommand{\pa}{\partial}

\newcommand{\rar}{\rightarrow}

\newcommand{\gsim}{ \lower .75ex \hbox{$\sim$} \llap{\raise .27ex \hbox{$>$}} }
\newcommand{\lsim}{ \lower .75ex \hbox{$\sim$} \llap{\raise .27ex \hbox{$<$}} }

\begin{document}

%\preprint{{\bf ***~\jobname.tex~***}}

\title{
Spectral probes of the holographic Fermi groundstate:\\
dialing between the electron star and AdS Dirac hair
\\[.5in]}

\author{Mihailo~\v{C}ubrovi\'{c}}
\author{Yan Liu}
\author{Koenraad Schalm}
\author{Ya-Wen Sun}
\author{Jan Zaanen}
\affiliation{
~\\
\mbox{Institute Lorentz for Theoretical Physics, Leiden University}\\
\mbox{P.O. Box 9506, Leiden 2300RA, The Netherlands}}

\thanks{Email addresses: {cubrovic, liu, kschalm, sun, jan@lorentz.leidenuniv.nl.}}

\begin{abstract}

\vspace*{.3in} We argue that the electron star and the AdS Dirac
hair solution are two limits of the free charged Fermi gas in AdS.
Spectral functions of holographic duals to probe fermions in the
background of electron stars have a free parameter that quantifies
the number of constituent fermions that make up the charge and
energy density characterizing the electron star solution. The
strict electron star limit takes this number to be infinite. The
Dirac hair solution is the limit where this number is unity. This
is evident in the behavior of the distribution of holographically
dual Fermi surfaces. As we decrease the number of constituents in a fixed electron star background
the number of Fermi surfaces also decreases. 
An improved holographic Fermi groundstate should be a configuration that shares the qualitative properties of both limits.
% We argue that an improved holographic Fermi groundstate that shares the qualitative properties of both limits remains very mysterious from a condensed matter perspective.
 % We speculate that in a dynamical theory dissipative RG-flow will naturally drive this number of constituents down and that the true holographic groundstate is the Dirac Hair configuration dual to a single Fermi Liquid.
\end{abstract}

\pacs{??.??}

\maketitle

\newpage
%%%%%%%%%%%

\def\be{\begin{equation}}
\def\ee{\end{equation}}
\def\bea{\begin{eqnarray}}
\def\eea{\end{eqnarray}}

\def\qeff{q_\mathtt{eff}}
\section{Introduction}

The insight provided by the application of the AdS/CFT
correspondence to finite density Fermi systems has given brand new
perspectives on the theoretical robustness of non-Fermi liquids
\cite{Liu:2009dm,Cubrovic:2009ye,Faulkner:2009wj}; on an
understanding of  the non-perturbative stability of the regular
Fermi liquid equivalent to order parameter universality for bosons
\cite{Cubrovic:2010bf,schalm2}, and most importantly on the notion
of fermionic criticality: Fermi systems with no scale. In essence
strongly coupled conformally invariant fermi systems are one
answer to the grand theoretical question of fermionic condensed
matter: {\em Are there finite density Fermi systems that do not
refer at any stage to an underlying perturbative Fermi gas?}

It is natural to ask to what extent AdS/CFT can provide
a more complete answer to this question. 
Assuming, almost tautologically,  that the
underlying system is strongly coupled and there is in addition
some notion of a large $N$ limit, the Fermi system is dual to
classical general relativity with a negative cosmological constant
coupled to charged fermions and electromagnetism. As AdS/CFT maps
quantum numbers to quantum numbers, finite density configurations
of the strongly coupled large $N$ system correspond to solutions
of this Einstein-Maxwell-Dirac theory with finite charge density.
Since the AdS  fermions are the only object carrying charge,
and the gravity system is weakly coupled, one is immediately
inclined to infer that the generic solution is a weakly coupled
charged Fermi gas coupled to AdS gravity: in other words an AdS
electron star \cite{Tong,Hartnoll:2010gu}, the charged equivalent
of a neutron star in asymptotically anti-de Sitter space
\cite{deBoer:2009wk,Arsiwalla:2010bt}.

Nothing can seem more straightforward. Given the total charge
density $Q$ of interest, one constructs the free fermionic
wavefunctions in this system, and fills them one by one in
increasing energy  until the total charge equals $Q$. For
macroscopic values of $Q$ these fermions themselves will backreact
on the geometry. One can compute this backreaction; it changes the
potential for the free fermions at subleading order. Correcting
the wavefunctions at this subleading order, one converges on the
true solution order by order in the gravitational strength
$\kappa^2 E^2_{full~system}$. Here $E_{full~system}$ is the energy carried by the Fermi system and $\kappa^2$ is the gravitational
coupling constant $\kappa^2=8\pi G_{Newton}$ in the AdS gravity
system. Perturbation theory in $\kappa$ is dual to the $1/N$ expansion in the associated condensed matter
system.

The starting point of the backreaction computation is to follow
Tolman-Oppenheimer-Volkov (TOV) and use a Thomas-Fermi (TF) approximation for
the lowest order one-loop contribution
\cite{deBoer:2009wk,Tong,Hartnoll:2010gu,Arsiwalla:2010bt}. The
Thomas-Fermi approximation applies when the number of constituent
fermions making up the Fermi gas is infinite. For neutral fermions
this equates to the statement that the energy-spacing between the
levels is neglible compared to the chemical potential associated
with $Q$, $\Delta E/\mu \rightarrow 0$. For charged fermions the
Thomas-Fermi limit is more direct: it is the limit $q/Q
\rightarrow 0$ where $q$ is the charge of each constituent
fermion. \footnote{For a fermion in an harmonic
oscillator potential $E_n=\hbar(n-1/2)\omega$: thus $\Delta
E/E_{total} = 1/\sum_{1}^N(n-1/2) =2/N^2.$}

% TOV=TFermi

% number of constituents

% each constuent has a wavefunction/Pauli Dirac

% TF is the collective effect
This has been the guiding principle behind the approaches
\cite{deBoer:2009wk,Tong,Arsiwalla:2010bt,Hartnoll:2010gu,Puletti:2010de,Hartnoll:2010ik} and the recent papers \cite{StelSpec,SLQL}, with
the natural assumption that all corrections beyond Thomas-Fermi
are small quantitative changes rather than qualitative ones. On
closer inspection, however, this completely natural TF-electron
star poses a number of puzzles. The most prominent perhaps arises
from the AdS/CFT correspondence finding that every normalizable
fermionic wavefunction in the gravitational bulk corresponds to a
fermionic quasiparticle excitation in the dual condensed matter system. In particular
occupying a particular wavefunction is dual to having a particular
Fermi-liquid state \cite{Cubrovic:2010bf}. In the Thomas Fermi
limit the gravity dual thus describes an infinity of Fermi
liquids, whereas the generic condensed matter expectation would
have been that a been that a single(/few) liquid(s) would be the
generic groundstate away from the strongly coupled fermionic
quantum critical point at zero charge density. This zoo of Fermi
surfaces is already present in the grand canonical approaches at
fixed $\mu$ (extremal AdS-Reissner-Nordstr\"{o}m (AdS-RN) black holes)
\cite{Faulkner:2009wj} and a natural explanation would be that
this is a large $N$ effect. This idea, that the gravity theory is
dual to a condensed matter system with $N$ species of fermions,
and increasing the charge density ``populates'' more and more
of the distinct species of Fermi liquids, is very surprising from the condensed matter perspective.
Away from criticality one would expect the generic groundstate to be a single Fermi-liquid or some broken state due to pairing.
To pose the puzzle sharply, once one has a fermionic quasiparticle one should be able to adiabatically continue it to a free Fermi gas, which would imply that the free limit of the strongly coupled fermionic CFT is not a single but  a system of order $N$ fermions with an ordered distribution of fermi-momenta.
A possible explanation of the multitude of Fermi surfaces that is consistent with a single Fermi surface at weak coupling is that AdS/CFT describes  so-called ``deconfined and/or fractionalized Fermi-liquids'' where the number of Fermi surfaces is directly tied to the coupling strength \cite{Hartnoll:2010xj,Sachdev:2010um,Huijse:2011hp,StelSpec,SLQL}. 
It would argue that fermionic quantum criticality goes hand in hand with fractionalization for which there is currently scant experimental evidence.

The second puzzle is more technical. Since quantum numbers in the
gravity system equal the quantum numbers in the dual condensed
matter system, one is inclined to infer that each subsequent AdS
fermion wavefunction has incrementally higher energy than the
previous one. Yet analyticity of the Dirac equation implies that
all normalizable wavefunctions must have strictly vanishing energy
\cite{HongFaulkner}. It poses the question how the order in which
the fermions populate the Fermi gas is determined.

The third puzzle is that in the Thomas Fermi limit the
Fermi gas is gravitationally 
strictly confined to a bounded region: famously, the
TOV-neutron star has an edge. In AdS/CFT, however, all information
about the dual condensed matter system is read off at asymptotic
AdS infinity. Qualitatively, one can think of AdS/CFT as an
``experiment'' analogous to probing a spatially confined Fermi gas
with a tunneling microscope held to the exterior of the trap.
Extracting the information of the dual condensed matter system is
probing the AdS Dirac system confined by a gravitoelectric trap
instead of a magneto-optical trap for cold atoms. Although the
Thomas-Fermi limit should reliably capture the charge and
energy densities in the system, its abrupt non-analytic change at
the edge (in a trapped system) and effective absence of a density
far away from the center are well known to cause qualitative
deficiencies in the description of the system. Specifically
Friedel oscillations --- quantum interference in the outside tails of the
charged fermion density, controlled by the ratio $q/Q$ and
measured by a tunneling microscope --- are absent. Analogously,
there could be qualitative features in the AdS asymptotics of both
the gravito-electric background and the Dirac wavefunctions in
that adjusted background that are missed by the TF-approximation.
The AdS asymptotics in turn {\em specify} the physics of the dual
condensed matter system and since our main interest is to use
AdS/CFT to understand quantum critical fermion systems where $q/Q$
is finite, the possibility of a qualitative change inherent in the
Thomas Fermi limit should be considered.

There is another candidate AdS description of the dual of a
strongly coupled finite density Fermi system: the AdS black hole
with Dirac hair \cite{Cubrovic:2010bf,schalm2}. One arrives at
this solution when one starts one's reasoning from the dual
condensed matter system, rather than the Dirac fields in AdS
gravity.  Insisting that the system collapses to a generic single
species Fermi-liquid ground state, the dual gravity description is
that of an AdS Einstein-Dirac-Maxwell system with a single nonzero
normalizable Dirac wavefunction. To have a macroscopic
backreaction the charge of this single Dirac field must be
macroscopic. The intuitive way to view this solution is as the
other simplest approximation to free Fermi gas coupled to gravity.
What we mean is that the full gravito-electric response is in all
cases  controlled by the total charge $Q$ of the
solution: as charge is conserved it is proportional to the
constituent charge $q$ times the number of fermions $
n_{F_{AdS}}$ and the two simple limits correspond to $n_F \rar
\infty, q\rar 0$ with $Q=qn_F$ fixed or $n_F\rar 1, q\rar Q$. The
former is the Thomas-Fermi electron star, the latter is the AdS
Dirac hair solution. In the context of AdS/CFT there is a
significant difference between the two solutions in that the Dirac
Hair solution clearly does not give rise to the puzzles 1, 2 and 3:
there is by construction no zoo of Fermi-surfaces and therefore no
ordering. Moreover since the wavefunction is demanded to be
normalizable, it manifestly encodes the properties of the system
at the AdS boundary.  On the
other hand the AdS Dirac hair solution does pose the puzzle that
under normal conditions the total charge $Q$ is much larger than
the constituent charge $q$ both from the gravity/string theory point of view and the condensed matter perspective. Generically one would expect a Fermi gas electron star rather than Dirac hair.

In this article we shall provide evidence for this  point of view
that the AdS electron star and the AdS Dirac hair solution are two
limits of the same underlying system. Specifically we
shall show that (1) the electron star solution indeed has the
constituent charge as a free parameter which is formally sent to
zero to obtain the Thomas-Fermi approximation. (2) The number of
normalizable wavefunctions in the electron star depend on the
value of the constituent charge $q$. We show this by computing the
electron star spectral functions. They depend in similar way on
$q$ as the first AdS/CFT Fermi system studies in an
AdS-RN background. In the formal limit where
$q\rar Q$, only one normalizable mode remains and the spectral
function wavefunction resembles the Dirac Hair solution,
underlining their underlying equivalence. Since
both approximations have qualitative differences % deficiencies
as a
% qualitative
description of the AdS dual to strongly coupled fermionic
systems, it argues that an improved approximation which has
characteristics of both is called for.

The results here are complimentary to and share an analysis of electron star spectral functions with the two recent articles \cite{StelSpec} and \cite{SLQL} that appeared in the course of this work (see also \cite{Iizuka:2011hg} for fermion spectral functions in general Lifshitz backgrounds). Our motivation to probe the system away from the direct electron star limit differs: we have therefore been more precise in defining this limit and in the analysis of the Dirac equation in the electron star background.

\section{Einstein-Maxwell theory coupled to charged fermions}

The Lagrangian that describes both the electron star and Dirac Hair approximation is Einstein-Maxwell theory coupled to charged matter
\begin{eqnarray}
\label{eq:01}
S &=& \int d^4x \sqrt{-g}\bigg[\frac{1}{2\kappa^2}\big(R+\frac{6}{L^2}\big)-\frac{1}{4q^2}F^2
+\mathcal{L}_{\mathtt{matter}}(e_{\mu}^A,A_{\mu})\bigg],
\end{eqnarray}
where $L$ is the AdS radius, $q$ is the electric charge and $\kappa$ is the gravitational coupling constant.
It is useful to scale the electromagnetic interaction to be of the same order as the gravitational interaction and measure all lengths in terms of the AdS radius $L$:
\begin{eqnarray}
  \label{eq:5}
  g_{\mu\nu} \rightarrow L^{2}{g}_{\mu\nu}~,~~A_{\mu} \rightarrow \frac{q L}{\kappa} {A}_{\mu}.
\end{eqnarray}
The system then becomes
\begin{eqnarray}
\label{eq:1}
S &=& \int d^4x \sqrt{-g}\bigg[\frac{L^2}{2\kappa^2}\bigg(R+6-\frac{1}{2}F^2\bigg)
+L^4\mathcal{L}_{\mathtt{matter}}(L{e}_{\mu}^A,\frac{qL}{\kap}A_{\mu})\bigg].
\end{eqnarray}
% with
% \begin{eqnarray}
% \mathcal{L}_{\mathtt{fermions}}(e_{\mu}^a,A_{\mu})=  -L^3\left(\bar{\Psi}e^{\mu}_A\Gamma^A(\partial_{\mu}+ \frac{1}{4} \ome_{\mu}^{BC}\Gamma_{BC}+ i\frac{qL}{\kappa} A_{\mu})-mL\right)\Psi
% \end{eqnarray}
% with equations of motion
% \begin{eqnarray}
%   \label{eq:3}
%    R_{\mu\nu}-\frac{1}{2}g_{\mu\nu}R - 3 g_{\mu\nu} &=& \left( F_{\mu\rho}F_{\nu}^{~~\rho}-\frac{1}{4}g_{\mu\nu} F_{\rho\sig}F^{\rho\sig} +   \kap^2 L^2 T_{\mu\nu}^{\mathtt{matter}}\right)  \non
% D_{\mu} F^{\mu\nu} = \kap^2L^2 J^{\nu}_{\mathtt{matter}}
% \end{eqnarray}
Note that in the rescaled variables the effective charge of charged matter now depends on the ratio of the electromagnetic to gravitational coupling constant: $q_\mathtt{eff} = qL/\kap$.
For the case of interest, charged fermions, the Lagrangian in these variables is
\begin{eqnarray}
\label{eq:16}
L^4\mathcal{L}_{\mathtt{fermions}}(Le_{\mu}^A,\frac{qL}{\kap}A_{\mu}) &=& % -\bar{\Psi}e^{\mu}_A\Gamma^A(\partial_{\mu}+ \frac{1}{4} \ome_{\mu}^{BC}\Gamma_{BC}+ i  A_{\mu})\Psi - m\bar{\Psi}\Psi \non
% \mathcal{L}_{\mathtt{fermions}}(e_{\mu}^a,A_{\mu})&=&
-\frac{L^2}{\kap^2}\bar{\Psi}\left[e^{\mu}_A\Gamma^A\big(\partial_{\mu}+ \frac{1}{4} \ome_{\mu}^{BC}\Gamma_{BC}- i\frac{qL}{\kappa} A_{\mu}\big)-mL\right]\Psi,
\end{eqnarray}
where $\bar{\Psi}$ is defined as $\bar{\Psi}=i\Psi^{\dagger}\Gamma^0$.
Compared to the conventional normalization the Dirac field has been made dimensionless $\Psi= \kap\sqrt{L}\psi_{\mathrm{conventional}}$. With this normalization all terms in the action have a factor $L^2/\kap^2$ and it will therefore scale out of the equations of motion
\begin{eqnarray}
  \label{eq:3}
   R_{\mu\nu}-\frac{1}{2}g_{\mu\nu}R - 3 g_{\mu\nu} &=& \left( F_{\mu\rho}F_{\nu}^{~~\rho}-\frac{1}{4}g_{\mu\nu} F_{\rho\sig}F^{\rho\sig} + T_{\mu\nu}^{\mathtt{fermions}}\right),  \non
D_{\mu} F^{\mu\nu} &=& - q_\mathtt{eff} J^{\nu}_{\mathtt{fermions}}
\end{eqnarray}
with 
\begin{eqnarray}
  \label{eq:2}
T_{\mu\nu}^{\mathtt{fermions}} &=& \frac{1}{2}\bar{\Psi}e_{A(\mu }\Gamma^A\bigg[\partial_{\nu)}+ \frac{1}{4} \ome_{\nu)}^{BC}\Gamma_{BC}- i\frac{qL}{\kap}  A_{\nu)}\bigg]\Psi - \frac{\kap^2L^2}{2} g_{\mu\nu} \mathcal{L}_{\mathtt{fermions}},  \\
J^{\nu}_{\mathtt{fermions}}&& = i\bar{\Psi}e^{\nu}_A\Gamma^A\Psi,
\end{eqnarray}
where the symmetrization is defined as $B_{(\mu}C_{\nu)}=B_{\mu}C_{\nu}+B_{\nu}C_{\mu}$
and the Dirac equation
\begin{eqnarray}
  \label{eq:17}
 \left[ e^{\mu}_A\Gamma^A\big(\partial_{\mu}+ \frac{1}{4} \ome_{\mu}^{BC}\Gamma_{BC}- i\frac{qL}{\kappa} A_{\mu}\big)-mL\right]\Psi =0.
\end{eqnarray}

The stress-tensor and current are to be evaluated in the specific state of the system. For a single excited wavefunction, obeying \eqref{eq:17}, this gives the AdS Dirac hair solution constructed in \cite{Cubrovic:2010bf}. (More specifically, the Dirac hair solution consists of a radially isotropic set of wavefunctions with identical momentum size $|\vec{k}|=\sqrt{k_x^2+k_y^2}$, such that the Pauli principle plays no role.)
For multiple occupied fermion states, even without backreaction due to gravity, adding the contributions of each separate solution to \eqref{eq:17} rapidly becomes very involved. In such a many-body-system, the collective effect of the multiple occupied fermion states is better captured in a ``fluid'' approximation
\be T_{\mu\nu}^{~\mathtt{fluid}}=(\rho+p)u_\mu u_\nu+pg_{\mu\nu},~~ ~~N_\mu^{~\mathtt{fluid}}=n u_\mu
\ee
with
\begin{eqnarray}
  \label{eq:6}
  \rho = \langle u^{\mu}T_{\mu\nu}u^{\nu} \rangle_{\mathtt{matter~only}}~,~~ n =-\langle u_{\mu}J^{\mu} \rangle_{\mathtt{matter~only}}.
\end{eqnarray}

In the center-of-mass rest frame of the multiple fermion system ($u_{\mu}=(e_{t\underline{0}},0,0,0)$), the expressions for the stress-tensor and charge density are given by the one-loop equal-time expectation values (as opposed to time-ordered correlation functions)
\begin{eqnarray}
  \label{eq:7}
  \rho = \langle \bar{\Psi}(t) e^{t}_{\underline{0}}\Gamma^{\underline{0}} (\partial_t+ \frac{1}{4}\ome_t^{AB}\Gamma_{AB}- i q_\mathtt{eff} A_t) \Psi(t)\rangle.
\end{eqnarray}
By the optical theorem the expectation value is equal to  twice imaginary part of the Feynman propagator\footnote{From unitarity for the $S$ matrix $S^{\dagger}S=1$ one obtains the optical theorem $T^{\dagger}T=2{\rm Im} T$ for the transition matrix $T$ defined as $S\equiv 1+iT$.}
\begin{eqnarray}
  \label{eq:8}
  \rho = \lim_{t\rar t'}2{\rm Im}{\rm Tr} \left[ e^{t}_{\underline{0}} \Gamma^{\underline{0}} (\partial_t+\frac{1}{4} \ome_t^{AB}\Gamma_{AB}- i q_\mathtt{eff} A_t)   G_F^{AdS}(t',t)\right].
\end{eqnarray}
In all situations of interest, all background fields will only have dependence on the radial AdS direction;
in that case the spin connection can be % ignored\footnote{In metrics with purely radial dependence such as AdS, the spin connection is actually
  absorbed in the normalization of the spinor wavefunction.\footnote{i.e. one can redefine spinors $\chi(r) = f(r)\Psi(r)$ such that the connection term is no
 longer present in the equation of motion.}
In an adiabatic approximation for the radial dependence of $e_{t\underline{0}}$ and $A_{t}$ --- where $\mu_{\mathtt{loc}}(r) = q_\mathtt{eff} e^t_{\underline{0}}(r) A_t(r)$  and $\ome(r) = -ie^t_{\underline{0}}(r)\partial_t$; % \comment{HERE I CHOSE I DEFINITE SIGN FOR $\ome$ THIS IS WHAT INTRODUCES THE Theta-FUNCTION}
--- this yields the known expression for a many-body-fermion system at finite chemical potential
% \begin{eqnarray}
%   \label{eq:9}
%   \rho = \lim_{\beta \rar \infty} \int d^2kdk_zd\ome (\ome(z) - \mu_{\mathtt{loc}}(z)) (1 +\tanh(\frac{\beta}{2}(\ome-\mu)) \delta(- \Gamma^0\ome+\Gamma^ik_i+\Gamma^zk_z+\Gamma^0\mu-im)
% \end{eqnarray}
%
\def\muloc{\mu_{\mathrm{loc}}}
\begin{eqnarray}
  \label{eq:9}
  \rho(r) &=& \lim_{\beta \rar \infty} 2 \int \frac{d^3kd\ome}{(2\pi)^4} \left[\ome(r) - \mu_{\mathtt{loc}}(r)\right] {\rm Im Tr}\, i\Gamma^{\underline{0}}G_F^{\beta} (\ome,k) \non
&=& \lim_{\beta \rar \infty} \int \frac{dk d\ome}{4 \pi^3} \left[k^2(\ome - \mu)\right] \left[\frac{1}{2} -\frac{1}{2}\tanh(\frac{\beta}{2} (\ome-\mu))\right] {\rm Tr}(i\Gamma^0)^2\frac{\kap^2}{L^2}\pi \delta((\ome-\mu)-\sqrt{k^2+(mL)^2}) \non
% &=&\lim_{\beta \rar \infty} \frac{4\pi}{(2\pi)^3} \int k^2dk d\ome (\ome(z) - \mu_{\mathtt{loc}}(z)) \frac{e^{\frac{\beta(\ome-\mu)}{2}}}{e^{\frac{\beta(\ome-\mu)}{2}}+e^{-\frac{\beta(\ome-\mu)}{2}}} \delta((\ome-\mu)-\sqrt{k^2+m^2})\\
&=&\lim_{\beta \rar \infty} \frac{\kap^2}{\pi^2L^2} \int d\ome f_{FD}(\beta(\ome-\mu)) \left[(\ome-\mu)^2-(mL)^2\right]\left[\ome - \mu\right] \frac{(\ome-\mu)\theta(\ome-\mu-mL)}{\sqrt{(\ome-\mu)^2-(mL)^2}} \non
&=& \frac{1}{\pi^2}\frac{\kap^2}{L^2} \int_{mL}^{\mu_{\mathrm{loc}}} d E E^2 \sqrt{E^2-(mL)^2}~.
\end{eqnarray}

The normalization $\kap^2/L^2$ follows from the unconventional normalization of the Dirac field in eq. \eqref{eq:16}.\footnote{One can see this readily by converting the dimensionless definition of $\rho$, eq~\eqref{eq:7}, to the standard dimension. Using capitals for dimensionless quantities and lower-case for dimensionful ones
\begin{eqnarray*}
\rho &\sim& \langle \Psi \pa_T \Psi\rangle %\non
%&\sim& \kap^2L \langle\psi \pa_T \psi\rangle \non
%&\sim&
\sim \kap^2L^2 \langle\psi \pa_t\psi\rangle %\non
%& \sim&
\sim
\kap^2L^2 \int_m^{\mu} d\eps \eps^2\sqrt{\eps^2-m^2}
%\non &\sim &
\sim \frac{\kap^2}{L^2} \int_{mL}^{\mu L} dE E^2\sqrt{E^2-(mL)^2}
\end{eqnarray*}
with $\mu L=\mu_\mathtt{loc}$ above.} 
Similarly
\begin{eqnarray}
  \label{eq:11}
  n =\frac{1}{\pi^2}\frac{\kap^2}{L^2} \int_{mL}^{\mu_{\mathrm{loc}}}d E E \sqrt{E^2-(mL)^2} = \frac{1}{3\pi^2}\frac{\kap^2}{L^2} ({\mu_{\mathrm{loc}}}^2-(mL)^2)^{3/2}.
\end{eqnarray}

The adiabatic approximation is valid for highly localized wavefunctions, i.e. the expression must be dominated by high momenta (especially in the radial direction). The exact expression on the other hand will not have a continuum of solutions to the harmonic condition $- \Gamma^0\ome+\Gamma^ik_i+\Gamma^zk_z-\Gamma^0\mu_{\mathrm{loc}}-imL =0$. Normalizable solutions to the AdS Dirac equations only occur at discrete momenta --- one can think of the gravitational background as a potential well. The adiabatic approximation is therefore equivalent to the Thomas-Fermi
approximation for a Fermi-gas in a box.

To get an estimate for the parameter range where the adiabatic approximation holds, consider the adiabatic bound $\pa_r \mu_{\mathtt{loc}}(r) \ll \mu_{\mathtt{loc}}(r)^2$. % (In all situations of interest, all background fields will only have dependence on the radial AdS diraction).
Using the field equation for $A_{\underline{0}} =\mu_\mathtt{loc}/ q_\mathtt{eff}$:
\begin{eqnarray}
\pa_r^2 \mu_\mathtt{loc}  \sim q^2_\mathtt{eff} n,
\end{eqnarray}
this bound is equivalent to requiring
\begin{eqnarray}
  \label{eq:12}
  \pa_r^2 \mu_\mathtt{loc}  \ll \pa_r \mu_\mathtt{loc} ^2 ~~~\Rightarrow ~~~ (\frac{qL}{\kap})^2 n \ll 2 \mu_\mathtt{loc}  \pa_r\mu_\mathtt{loc}  ~~~\Rightarrow~~~  (\frac{qL}{\kap})^2 n \ll \mu_\mathtt{loc}^3
\end{eqnarray}
where in the last line we used the original bound again.
If the chemical potential scale is considerably higher than the mass of the fermion, we may use \eqref{eq:11} to approximate $n  \sim \frac{\kap^2}{L^2}{\mu_{\mathtt{loc}}^3}$. Thus the adiabatic bound is equivalent to,
\begin{eqnarray}
  \label{eq:4}
  q = \frac{\qeff\kap}{L}\ll 1
\end{eqnarray}
the statement that the constituent charge of the fermions is infinitesimal.
Note that in the rescaled action (\ref{eq:1}, \ref{eq:16}), $L/\kap$ plays the role of $1/\hbar$, and eq. \eqref{eq:4} is thus equivalent to the semiclassical limit $\hbar \rar 0$ with $\qeff$ fixed. Since AdS/CFT relates $L/\kappa \sim N_c$ %$\kap/L \sim g^2_{YM}N_c$
this acquires the meaning in the context of holography that there is a large $N_c$ scaling limit \cite{StelSpec,SLQL} of the CFT with fermionic operators where the RG-flow is ``adiabatic''. Returning to the gravitational description
 the additional assumption that the chemical potential is much larger than the mass is equivalent to 
\begin{eqnarray}
  \label{eq:15} \frac{Q^{\mathtt{total}}_{\mathtt{phys}}}{V_{\mathtt{spatial~AdS}}}=\frac{LQ^{\mathtt{total}}_{\mathtt{eff}}}{\kappa V_{\mathtt{spatial~AdS}}} &\equiv& \frac{L}{ \kappa V_{\mathtt{spatial~AdS}}} \int dr \sqrt{-g_{\rm induced}}\left( \qeff n \right)\\ \nonumber
  &\simeq& \frac{1}{ V_{\mathtt{spatial~AdS}}} \int dr \sqrt{-g}  \frac{\qeff\kappa}{L} \mu_{\mathtt{loc}}^3(r) ~~~\gg~~~ q  (mL)^3~.
\end{eqnarray}
This implies that the total charge density in AdS is much larger than that of a single charged particle (as long as  $mL \sim 1$). The adiabatic limit is therefore equivalent to a thermodynamic limit where the Fermi gas consists of an infinite number of constituents, $n\rar \infty$, $q \rar 0$ such that the total charge $Q \sim nq$ remains finite.

The adiabatic limit of a many-body fermion system coupled to gravity are the Tolman-Oppenheimer-Volkov
equations. Solving this in asymptotically AdS gives us the charged neutron or electron star constructed in \cite{Hartnoll:2010gu}. Knowing the quantitative form of the adiabatic limit, it is now easy to distinguish the electron star solution from the ``single wavefunction'' Dirac Hair solution. The latter is trivially the single particle limit $n\rar 1$, $q \rar Q$ with the total charge $Q$ finite. The electron star and Dirac Hair black hole are opposing limit-solutions of the same system. We shall now make this connection more visible by identifying a formal dialing parameter that interpolates between the two solutions.
% Integrating once
% \begin{eqnarray}
%   \label{eq:13}
%   \pa_z A_t = \frac{q\kappa}{L} \int dz n
% \end{eqnarray}
% The left hand side equals $\pa_r \mu_{loc}$, whereas the righthand side equals the total charge density $N = \int dz n$. In equilibrium it equals $N \sim \mu^3$, and thus
% \begin{eqnarray}
%   \label{eq:14}
%   \frac{1}{\mu^3} \pa_r \mu = \frac{q\kappa}{L}g_eff^4 = g_eff^5 \left(\frac{\kappa}{L}\right)^{3/2} = \frac{q^5L}{\kappa}
% \end{eqnarray}

% Rescaling $m= \frac{q}{\kappa} \hat{m}$ one finds that
% \begin{eqnarray}
%   \label{eq:10}
%   \rho = \frac{1}{(2\pi)^2} \frac{q^4}{\kappa^4} \int^{\mu}_{\hat{m}} \eps^2 \sqrt{\eps^2-\hat{m}^2} d\eps
% \end{eqnarray}
% \def\geff{g_{\mathtt\mathtt{eff}}}
% Defining the dimensionless effective coupling $\geff^{4} = \frac{q^4 L^2}{\kappa^2}\frac{1}{(2\pi)^2}$ ($\geff^4=\beta$ in \cite{Hartnoll:2010gu})
% The fermion Lagrangian then becomes
% \begin{eqnarray}
% \mathcal{L}_{\mathtt{fermions}}(e_{\mu}^a,A_{\mu})=  -L^3\left(\bar{\Psi}e^{\mu}_A\Gamma^A(\partial_{\mu}+ \frac{1}{4} \ome_{\mu}^{BC}\Gamma_{BC}+ i\geff\sqrt{\frac{2\pi L}{\kappa}} A_{\mu})-\geff\sqrt{\frac{2\pi L}{\kappa}} \hat{m}\right)\Psi
% \end{eqnarray}
%-----------------------------------------------------

To do so we shall need the full adiabatic Tolman-Oppenheimer-Volkov
equations for the AdS electron star \cite{Hartnoll:2010gu}.
Since the fluid is homogeneous and isotropic, the background metric and electrostatic potential will respect these symmetries and will be of the form
(recall that we are already using ``dimensionless'' lengths, eq. \eqref{eq:5})
\be
ds^2=-f(r)dt^2+g(r)dr^2+r^2(dx^2+dy^2),~~~A=h(r)dt,
\ee
where $f(r),g(r),h(r)$ are functions of $r$; the horizon is located at $r=0$ and the boundary is at $r=\infty$.
Combining this ansatz  with a rescaling
% $\mu_\mathtt{loc} = q_\mathtt{eff} \hat{\mu} = q_{\mathtt{eff}}\frac{h(r)}{\sqrt{f(r)}$,
$mL= {q_\mathtt{eff}} \hat{m}$
  the bosonic background equations of motion become
\cite{Hartnoll:2010gu}
\begin{eqnarray} \label{eombg1}
\frac{1}{r}\bigg(\frac{f'}{f}+\frac{g'}{g}\bigg)-\frac{gh{\sig}}{\sqrt{f}}=&0,&~~~~~~~~~~~~~~ {\rho} = \frac{q_{\mathtt{eff}}^4\kap^2}{\pi^2L^2} \int_{\hat{m}}^{\frac{h}{\sqrt{f}}} d\eps \eps^2\sqrt{\eps^2 -\hat{m}^2}~, %\frac{1}{q_\mathtt{eff}^4}\rho
\non
%\label{eombg2}
\frac{f'}{rf}+\frac{h'^{2}}{2f}-g(3+{p})+\frac{1}{r^2}=&0, &~~~~~~~~~~~~~~%\hat{\sig}(\hat{\mu},\hat{m}) = q_\mathtt{eff} n
{\sig} = \frac{q_{\mathtt{eff}}^4\kap^2}{\pi^2L^2} \int_{\hat{m}}^{\frac{h}{\sqrt{f}}} d\eps \eps\sqrt{\eps^2 -\hat{m}^2}~,
\non
%\label{eombg3}
h''+\frac{2}{r}h'-\frac{g{\sig}}{\sqrt{f}}\bigg(\frac{rhh'}{2}+f\bigg)=&0, &~~~~~~~~~~~~~~ -{p}= {\rho}-\frac{h}{\sqrt{f}}{\sig}~,
\end{eqnarray}
where we have used that $\mu_\mathtt{loc}= q_\mathtt{eff}
h/\sqrt{f}$ and $\sigma=nq_\mathtt{eff}$ is the rescaled local charge density. What one immediately notes is that the Tolman-Oppenheimer-Volkov equations
of motion for the background only depend on the parameters
$\hat{\beta} \equiv
\frac{q_\mathtt{eff}^4\kap^2}{\pi^2L^2}$ and
$\hat{m}$, whereas the original Lagrangian and the fermion
equation of motion also depend on $q_\mathtt{eff} =
\left(\frac{\pi^2L^2\hat{\beta}}{\kap^2}\right)^{1/4}$. It is
therefore natural to guess that the parameter $\qeff=qL/\kappa$ will be
the interpolating parameter away from the adiabatic electron star
limit towards the Dirac Hair BH.
%\comment{Dialing paragraph}
%This is clearly the ratio of the total to the microscopic charge.

Indeed in these natural electron star variables the adiabatic bound \eqref{eq:4} translates into
\begin{eqnarray}
  \label{eq:18}
  \hat{\beta} \ll \frac{L^2}{\kap^2} = \frac{q^2_\mathtt{eff}}{q^2}~.
\end{eqnarray}
Thus we see that for a given electron star background with $\hat{\beta}$ fixed decreasing $\kap/L$  improves the adiabatic fluid  approximation whereas increasing $\kap/L$ makes the adiabatic approximation poorer and poorer.
``Dialing  $\kap/L$ up/down'' therefore {\em interpolates}  between the electron star and the Dirac Hair BH.  Counterintuively improving adiabaticity
by decreasing $\kap/L$ corresponds to increasing $\qeff$ for fixed $q$, but this is just a consequence of recasting the system in natural electron star variables.
A better way to view improving adiabaticity is to decrease the microscopic charge $q$ but while keeping $q_\mathtt{eff}$ fixed; this shows that a better way to think of $\qeff$ is as the total charge rather than the efffective constituent charge.

The parameter $\kap/L=q/\qeff$ parametrizes the gravitational coupling strength in units of the AdS curvature, and one might worry that ``dialing $\kap/L$ up'' pushes one outside the regime of classical gravity. This is not the case. One can easily have $\hat{\beta} \gg 1$ and tune $\kap/L$ towards or away from the adiabatic limit within the regime of classical gravity. From eq. \eqref{eq:4} we see that the edge of validity of the adiabatic regime $\hat{\beta} \simeq L^2/\kap^2$ is simply equivalent to a microscopic charge $q=1$ which clearly has a classical gravity description. It is not hard to see that the statement above is the equivalent of changing the level splitting in the Fermi gas, while keeping the overall energy/charge fixed.  In a Fermi gas microscopically both the overall energy and the level splitting depends on $\hbar$. Naively increasing $\hbar$ increases both, but one can move away from the adiabatic limit either by decreasing the overall charge density, keeping $\hbar$ fixed or by keeping the charge density fixed and raising $\hbar$. Using again the analogy between $\kap/L$ and $\hbar$, the electron star situation is qualitatively the same where one should think of $\hat{\beta} \sim q^4L^2/\kap^2$ parametrizing the microscopic charge. % \footnote{To apply the exact analogy one can keep the total energy $\rho \sim \beta$ fixed in units of the gravitational coupling while increasing/decreasing $\kap/L$, i.e. the combination $E_{\mathtt{total}}\kap \propto \rho\kap^4/L^4$. To do so $\beta $ }
One can either insist on keeping $\kap/L$ fixed and {\em increase} the microscopic charge $\hat{\beta}$ to increase the level splitting or one can keep $\hat{\beta}$ fixed and increase $\kap/L$. In the electron star, however, the background geometry changes with $\hat{\beta}$ in addition to the level splitting, and it is therefore more straightforward to keep $\hat{\beta}$ and the geometry fixed, while dialing $\kap/L$.

We will now give  evidence for our claim that the electron
star and Dirac Hair solution are two opposing limits. To do so, we
need to identify an observable that goes either beyond the
adiabatic background approximation or beyond the single particle
approximation. Since the generic intermediate state is still a
many-body fermion system, the more natural starting point is the
electron star background and perturb away from there. Realizing
then that the fermion equation of motion already depends directly
on the dialing parameter $\qeff$ the obvious observables are the
single fermion spectral functions in the electron star background.
Since one must specify a value for $\qeff$ to compute these, they
directly probe the microscopic charge of the fermion and are thus
always beyond the strict electron star limit $q\rightarrow 0$. In the
next two sections we will compute these and show that they indeed
reflect the interpretation of $\qeff$ as the  interpolating
parameter between the electron star and Dirac Hair BH.

%%%%%%%%%%%%%%%%%%%%%%%%%%%%%%%%%%%%%%
%%%%%%%%%%%%%%%%%%%%%%%%%%%%%%%%%%%%%%

\section{Fermion spectral functions in the electron star background}

%%%%%%%%%%%%%%%%%%%%%%%%%%%%%%%%%%%%%%
%%%%%%%%%%%%%%%%%%%%%%%%%%%%%%%%%%%%%%

To compute the fermion spectral functions in the electron star
background we shall choose a specific representative of the family
of electron stars parametrized by $\hat{\beta}$ and $\hat{m}$.
Rather than using $\hat{\beta}$ and $\hat{m}$ the metric of an
electron star is more conveniently characterized by its
Lifshitz-scaling behavior near the interior horizon $r \rar 0$.
From the field equations \eqref{eombg1} the limiting interior
behavior of $f(r), g(r), h(r)$ is
\begin{eqnarray}
  \label{eq:19}
  f(r) = r^{2z} + \ldots~,~~g(r) = \frac{g_{\infty}}{r^2}+\ldots~,~~h(r)={h_{\infty}}{r^z}+\ldots
\end{eqnarray}
The scaling behavior is determined by the dynamical critical
exponent $z$, which is a %known
function of $\hat{\beta},~\hat{m}$
\cite{Hartnoll:2010gu} and it is conventionally used to classify
the metric instead of $\hat{\beta}$. The full electron star metric
is then generated from this horizon scaling behavior by
integrating up an irrelevant RG-flow 
\cite{GubserNellore,Goldstein:2009cv}
\begin{eqnarray}\label{persol} f={r^{2z}}\bigg(1+f_1r^{-\alpha}+\dots\bigg),
 ~~g=\frac{g_\infty}{r^{2}}\bigg(1+g_1r^{-\alpha}+\dots\bigg),
 ~~h={h_\infty}{r^{z}}\bigg(1+h_1r^{-\alpha}+\dots\bigg).
\end{eqnarray}
with
\begin{eqnarray}
  \label{eq:20}
  \alp=\frac{2+z}{2} - \frac{\sqrt{9z^3-21z^2+40z-28-\hat{m}^2z(4-3z)^2}}{2\sqrt{(1-\hat{m}^2)z-1}}.
\end{eqnarray}
Scaling $f_1\to bf_1$ is equal to a coordinate transformation $r\to
b^{1/\alpha}r$ and $t\to b^{z/\alpha} t$, and the sign of $f_1$ is
fixed to be negative in order to be able to match onto an
asymptotically AdS$_4$ solution. Thus $f_1=-1$ and $g_1$ and $h_1$
are then uniquely determined by the equations of motion.

Famously, integrating the equations of motion up the RG-flow
outwards towards the boundary fails at a finite distance $r_s$.
This is the edge of the electron star. Beyond the edge of the
electron star, there is no fluid present and the spacetime is that
of an AdS$_4$-RN black hole with the metric
\begin{equation}
\label{eq:23} f={c^2}{r^2}-\frac{\hat{M}}{r}+\frac{\hat{Q}^2}{2r^2},~~~g=\frac{c^2}{f},~~~h=\hat{\mu}-\frac{\hat{Q}}{r}.
\end{equation}
Demanding the full metric is smooth at the radius of electron star
$r_s$ determines the constants $c,~\hat{M}$ and $\hat{Q}$. The dual field theory
is defined on the plane $ds^2=-c^2dt^2+dx^2+dy^2$.

\begin{figure}[ht]
\includegraphics[width=0.5\textwidth]{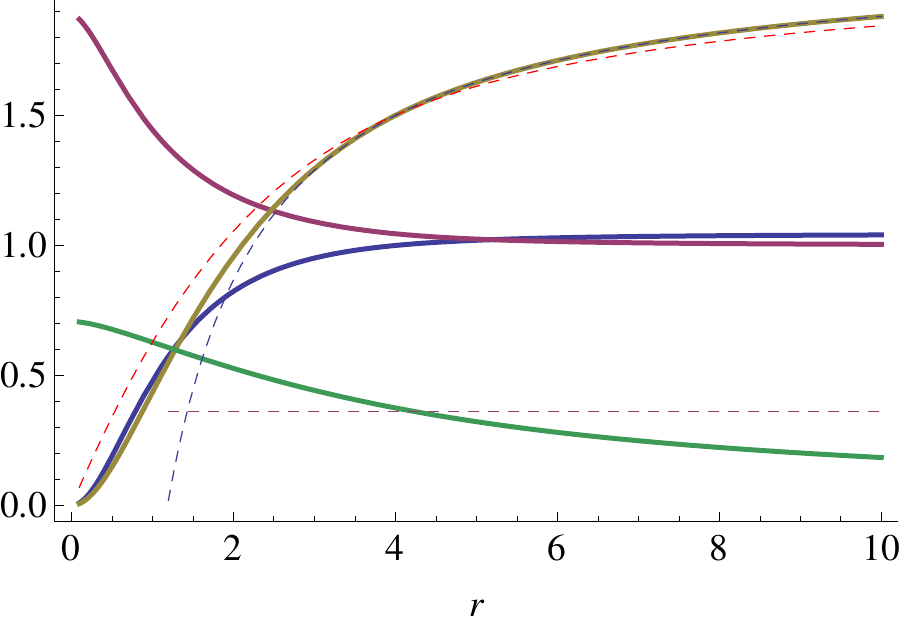}
\caption{\label{esbg}
Electron star metric for $z=2,\hat{m}=0.36$, $c\simeq1.021, \hat{M}\simeq3.601, \hat{Q}\simeq2.534, \hat{\mu}\simeq 2.132$ compared to pure AdS. Shown are $f(r)/r^2$ (Blue), $r^2 g(r)$ (Red) and $h(r)$ (Orange). The asymptotic AdS-RN value of $h(r)$ is the dashed blue line. For future use we have also given $\mu_{\mathtt{loc}} = h/\sqrt{f}$ (Green) and $\mu_{\qeff}= \sqrt{g^{ii}} h/\sqrt{f}$ (Red Dashed) At the edge of the star $r_s \simeq4.253$ (the intersection of the purple dashed line setting the value of $m_{\mathtt{eff}}$ with $\mu_{\mathtt{loc}}$) one sees the convergence to pure AdS in the constant asymptotes of $f(r)/r^2$ and $r^2 g(r)$.}
\end{figure}

The specific electron star background we shall choose without loss of generality is the one with $z=2,\hat{m}=0.36$ (Fig.
\ref{esbg})\footnote{This
background has $c\simeq1.021, \hat{M}\simeq3.601,
\hat{Q}\simeq2.534, \hat{\mu}\simeq 2.132$, $\hat{\beta}\simeq19.951,
~g_\infty\simeq1.887,~h_\infty=1/\sqrt{2},~ \alpha\simeq-1.626,~ f_1=-1,~
g_1\simeq-0.4457,~h_1\simeq-0.6445$.}, smoothly matched at $r_s\simeq4.25252$
onto a AdS-RN black-hole.

%%%%%%%%%%%%%%%%%%%%%%%%%%%%%%%%%%%%%%%
%%%%%%%%%%%%%%%%%%%%%%%%%%%%%%%%%%%%%%

 \bigskip
\def\meff{m_{\mathtt{eff}}}

The CFT fermion spectral functions now follow from solving the
Dirac equation in this background
\cite{Liu:2009dm,Cubrovic:2009ye}
\begin{eqnarray}
\label{Dirac}
\left[e^{\mu}_A\Gamma^A\left(\partial_\mu
+\frac{1}{4}\omega_{\mu AB}\Gamma^{AB}-i\qeff A_\mu\right)-\meff\right]\Psi=0
\end{eqnarray}
where $\qeff$ and $\meff$ in terms of the parameters of the electron
star equal
\bea
\qeff &=&\left(\frac{\pi^2L^2\hat{\beta}}{\kap^2}\right)^{1/4},~~~\meff
%&=&mL
=\qeff \hat{m}=\hat{m}\left(\frac{\pi^2L^2\hat{\beta}}{\kap^2}\right)^{1/4}.
\eea
For a given electron star background, i.e. a fixed $\hat{\beta}$,
$\hat{m}$ the fermion spectral function will therefore depend on
the ratio $L/\kap$.  For $L/\kap \gg \hat{\beta}^{1/2}$ the poles
in these spectral functions characterize the occupied states in a
many-body gravitational Fermi system that is well approximated
by the electron star. As $L/\kap$ is lowered for fixed
$\hat{\beta}$ the electron star background becomes a poorer and
poorer approximation to the true state and we should see this
reflected in both the number of poles in the spectral function and
their location. 

Projecting the Dirac equation onto two-component $\Gamma^{\underline{r}}$ eigenspinors
\begin{eqnarray}
\label{eq:21} \Psi_\pm=(-gg^{rr})^{-\frac{1}{4}}e^{-i\ome t+ik_ix^i}\left(\begin{array}{c}
y_{\pm}  \\
z_{\pm}
\end{array}\right)
\end{eqnarray}
and using isotropy to set $k_y=0$, one can choose a basis of Dirac matrices where one obtains two decoupled sets of two simple coupled equations \cite{Liu:2009dm}
\bea
\label{eq:49}
\sqrt{g_{ii}g^{rr}}(\partial_r\mp \meff\sqrt{g_{rr}})y_{\pm}&=& \mp i(k_x-u)z_{\mp}, \\
\label{eq: dirac2com}
\sqrt{g_{ii}g^{rr}}(\partial_r\pm \meff\sqrt{g_{rr}})z_{\mp}&=& \pm i(k_x+u)y_{\pm}
\eea
where $u=\sqrt{\frac{g_{ii}}{-g_{tt}}}(\ome+q_{\mathrm{eff}}h)$.
In this basis of Dirac matrices the CFT  Green's function $G=\langle
\bar\cO_{\psi_+} i\gamma^0\cO_{\psi_+}\rangle$ equals
\begin{eqnarray}
 G=\lim_{\epsilon\to 0}\epsilon^{-2mL}
\left(
\begin{array}{cc}
\xi_{+} &  0   \\
0  &   \xi_{-}
\end{array}
\right)\bigg|_{r=\frac{1}{\epsilon}},~~{\mathrm{where}}~~\xi_{+}=\frac{iy_-}{z_{+}},~~~ \xi_{-}=-\frac{iz_-}{y_{+}}.
\end{eqnarray}
Rather than solving the coupled equations (\ref{eq:49}) it is convenient to solve for $\xi_{\pm}$ directly \cite{Liu:2009dm},
\begin{eqnarray}
\label{eomxi} \sqrt{\frac{g_{ii}}{g_{rr}}}\partial_r\xi_{\pm}=-2\meff\sqrt{g_{ii}}\xi_{\pm}\mp(k_x\mp u)\pm(k_x\pm u)\xi_{\pm}^2.
\end{eqnarray}

For the spectral function $A={\rm Im Tr}G_{R}$ we are interested in the retarded Green function. This is obtained by imposing in-falling boundary conditions near the horizon $r=0$. Since the electron star is a ``zero-temperature'' solution this requires a more careful analysis than for a generic horizon. To ensure that the numerical integration we shall perform to obtain the full spectral function has the right infalling boundary conditions, we first solve eq.~(\ref{eomxi}) to first subleading order around $r=0$. There are two distinct branches. When $\ome\neq 0$ and $k_xr/\ome, r^2/\ome$  is small, the in-falling boundary condition near the horizon $r=0$ is (for $z=2$)% (at exactly $r=0$, only the first term in the following expression is needed but in practical calculations we have to start from a small but finite $r$ so the following expansion near $r=0$ is needed)
\bea \xi_{+}(r)&=&i-i\frac{k_xr}{\ome}+i\frac{(k_x^2-2i\meff \ome)r^2}{2\ome^2}-i\frac{f_1k_xr^{1-\alpha}}{2\ome}+\dots\nonumber\\
\xi_{-}(r)&=&i+i\frac{k_xr}{\ome}+i\frac{(k_x^2-2i\meff \ome)r^2}{2\ome^2}+i\frac{f_1k_xr^{1-\alpha}}{2\ome}+\dots.\eea
When $\ome=0$, i.e. $k_xr/\ome$ is large,
 and $r/k_x\to 0$,
\bea \label{bc2} \xi_{+}(r)&=&-1+\frac{(q_{\mathrm{eff}}h_{\infty}+\meff)r}{k_x}+\bigg(\frac{\ome}{k_xr}-\frac{\ome}{2\sqrt{g_{\infty}}k_x^2}\bigg)+\dots\nonumber\\
\xi_{-}(r)&=&1+\frac{(q_{\mathrm{eff}}h_{\infty}-\meff)r}{k_x}+\bigg(\frac{\ome}{k_xr}
-\frac{\ome}{2\sqrt{g_{\infty}}k_x^2}\bigg)+\dots, \eea
% Note that when $w=0$, we should impose the boundary conditions (\ref{bc2}). Different from the extremal RN-AdS black hole case \cite{Liu:2009dm}, for arbitrary $k_x$ here,
the boundary conditions (\ref{bc2}) become real.  As (\ref{eomxi})
are real equations, the spectral function vanishes in this case.
This is essentially the statement that all poles in the Green's
function occur at $\omega=0$ \cite{HongFaulkner}. Note that the fact that the electron star $\ome=0$ boundary conditions (\ref{eomxi}) are real for all values of $k$ is qualitatively different from the AdS-RN $\ome=0$ boundary conditions (eq. (26) in \cite{Liu:2009dm}). In the AdS-RN ``quantum-critical'' infrared governed by the near horizon AdS$_2\times\RR_2$ geometry, in general 
there is a special scale $k_o$ below which the boundary condition turns complex. This scale $k_o$ is related to the surprising existence of an oscillatory region in the spectral function.
One therefore infers that in a scale-ful Lifshitz infrared this oscillatory region is no longer present \cite{Guarrera:2011my,SLQL}. We will confirm this in section \ref{sec:fermi-surf-order}.

\subsection{Numerical results and discussion}

%%%%%%%%%%%%%%%%%%%%%%%%%%%%%%%%%%%%%%
\begin{figure}[t]
\includegraphics[width=0.60\textwidth]{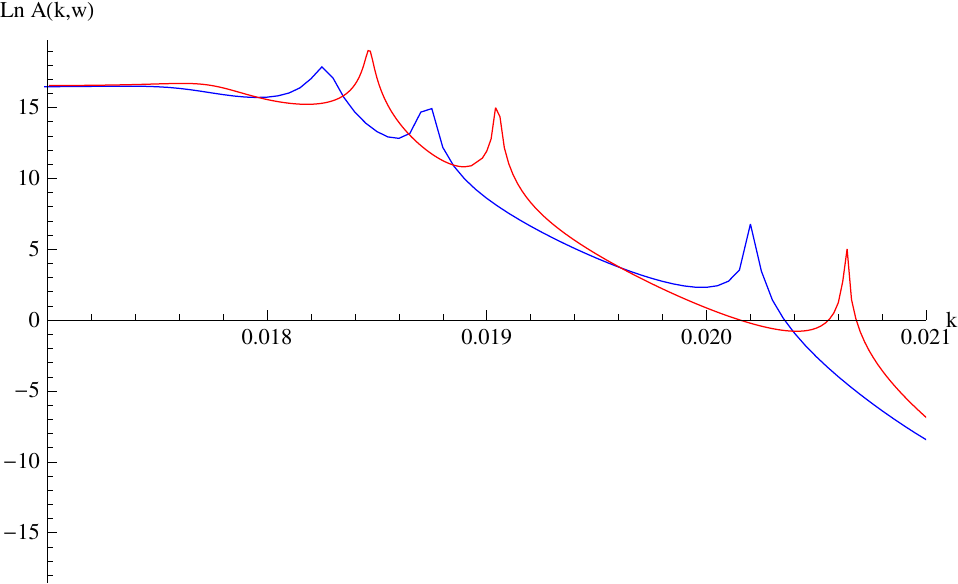}
\caption{\label{fig1} Electron star MDF spectral functions with multiple peaks as a function of $k$ for $\ome=10^{-5}, z=2, \hat{m}=0.36$. The blue curve is for $\kappa=0.091$; the red curve is for $\kappa=0.090$. 
Note that the vertical axis is logarithmic. Visible is the rapidly decreasing spectral weight and increasingly narrower width for each successive peak as $k_F$ increases.}
\end{figure}
%%%%%%%%%%%%%%%%%%%%%%%%%%%%%%%%%%%%%%

\begin{figure}[t]
(A)
\includegraphics[width=0.45\textwidth]{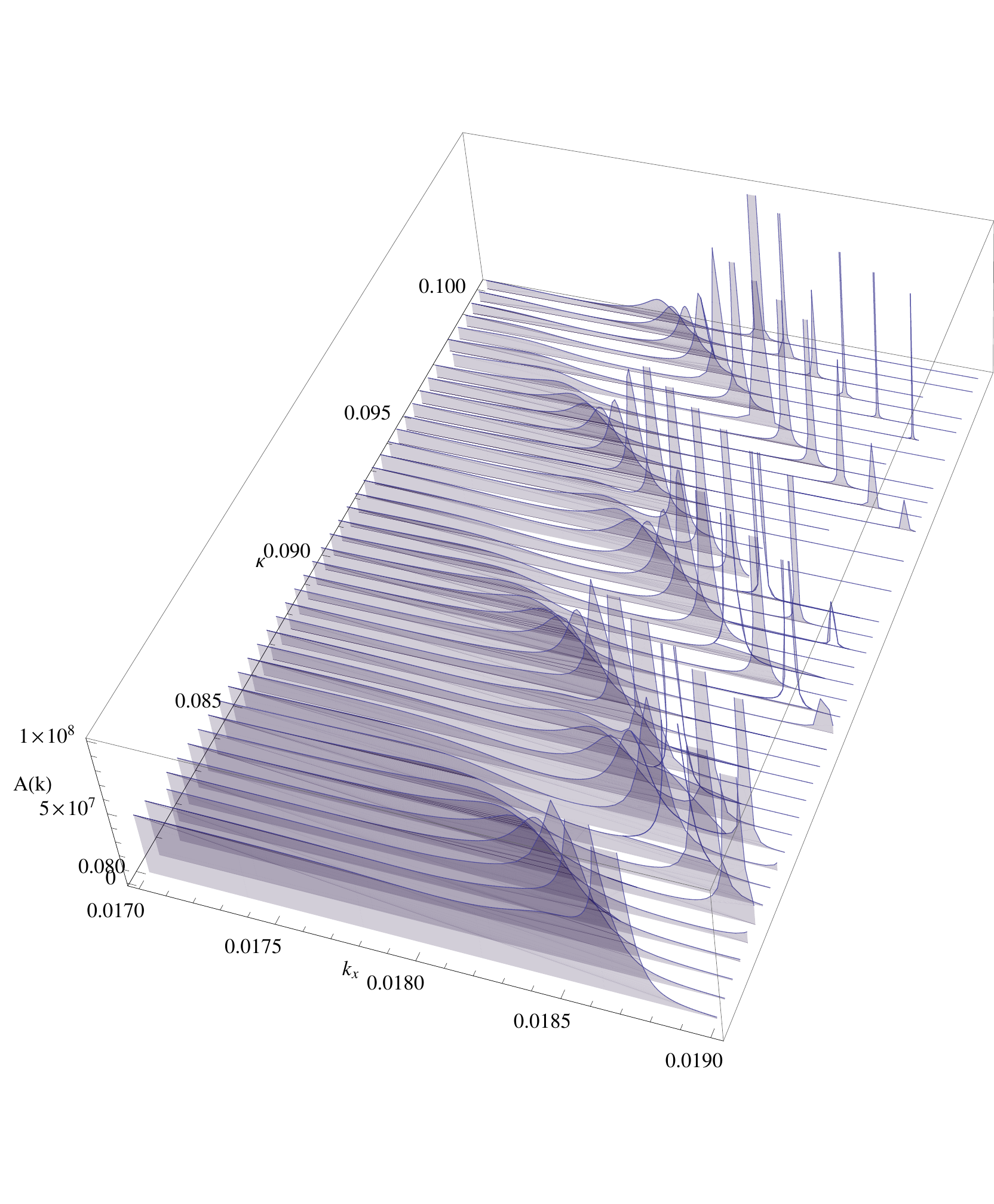}
%\hspace*{-0.42\textwidth}
%\includegraphics[width=0.45\textwidth]{ES-3DSpectra-19}
%\hspace*{-.1\textwidth}
\raisebox{-0.2in}{
\includegraphics[width=0.45\textwidth]{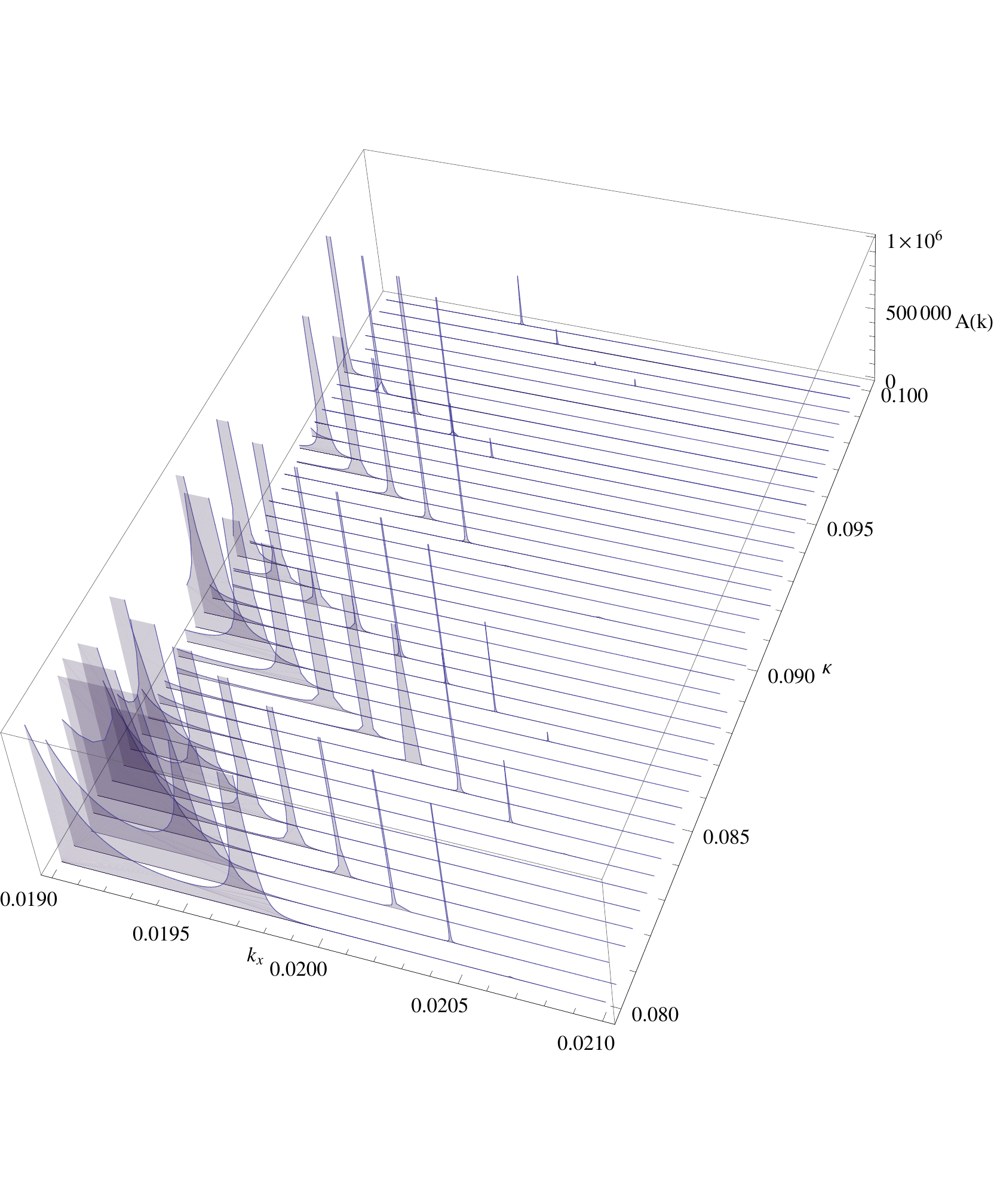}}

(B1)
%~\parbox[t]{0.35\textwidth}{
\includegraphics[width=0.4\textwidth]{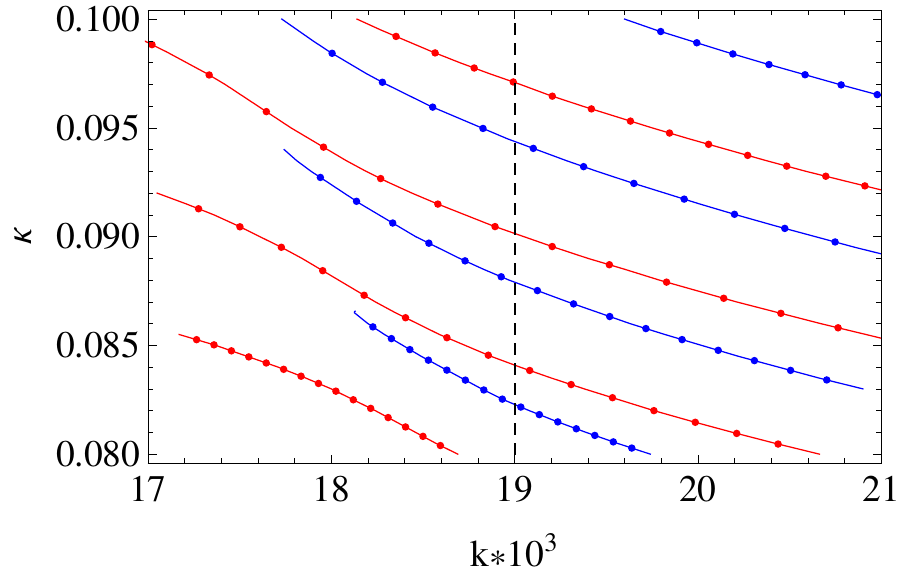}
(B2)
\includegraphics[width=0.4\textwidth]{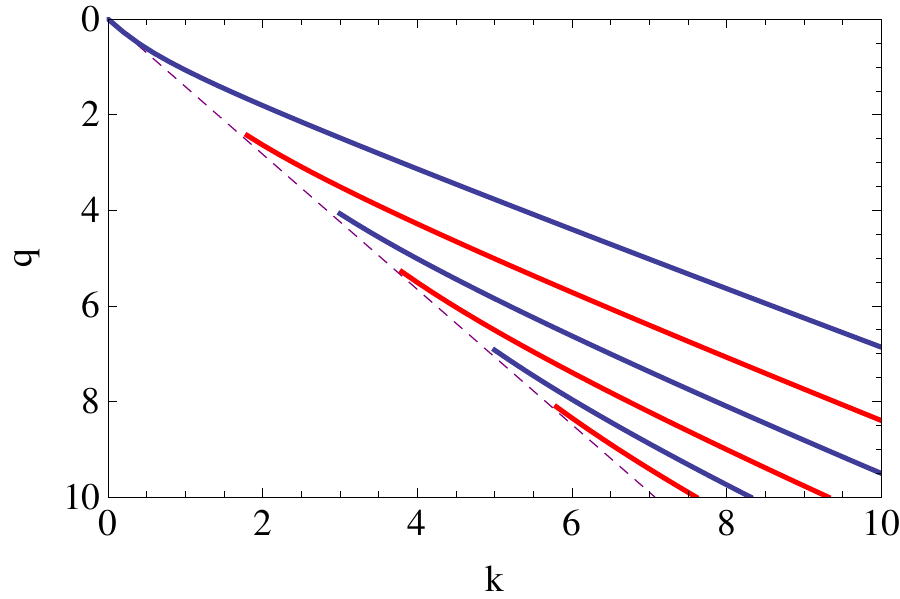}
%}
\caption{(A) Electron-star MDF spectral functions as a function of $\kappa$ for $z=2, \hat{m}=0.36, \ome=10^{-5}$. Because the peak height and weights decrease exponentially, we present the adjacent ranges $k\in [0.017,0.019]$ and $k\in[0.019,0.021]$ in two different plots with different vertical scale. (B1/B2) Locations of peaks of spectral functions as a function of $\kappa$: comparison between the electron star (B1) for $z=2, \hat{m}=0.36, \ome=10^{-5}$
(the dashed gray line denotes the artificial separation in the 3D representations in (A)) and AdS-RN (B2) for $m=0$ as a function of $q$ in units where $\mu=\sqrt{3}$  These two Fermi-surface `spectra' are qualitatively similar. \label{esspec}}
\end{figure}

We can now solve for the spectral functions numerically. In Fig. \ref{esspec}  we plot the momentum-distribution-function (MDF) (the spectral function as a function of $k$)
for fixed $\ome=10^{-5},~z=2,~\hat{m}=0.36$ while changing the value of $\kappa$. Before we comment on the dependence on $\qeff \sim  \kappa^{-1/2}$ which studies the deviation away from the adiabatic limit of a given electron star background ({i.e. fixed dimensionless  charge and fixed dimensionless energy density}), there are several striking features that are immediately apparent:
\begin{itemize}
\item As expected, there is a multitude of Fermi surfaces. They have very narrow width and their spectral weight decreases rapidly for each higher Fermi-momentum $k_F$ (Fig. \ref{fig1}). This agrees with the exponential width $\Gam \sim \exp( - \left(\frac{k^{z}}{\ome}\right)^{1/(z-1)})$  predicted by \cite{Faulkner:2010tq} for gravitational backgrounds that are Lifshitz in the deep interior, which is the case for the electron star. This prediction is confirmed in \cite{Iizuka:2011hg,StelSpec,SLQL} and the latter two articles also show that the weight decreases in a corresponding exponential fashion. This exponential reduction of both the width and the weight as $k_F$ increases explains why we only see a finite number of peaks, though we expect a very large number. In the next section we will be able to count the number of peaks, even though we cannot resolve them all numerically.
\item The generic value of $k_F$ of the peaks with visible spectral weight is {\em much} smaller than the effective chemical potential $\mu$ in the boundary field theory. This is quite different from the RN-AdS case where the Fermi momentum and chemical potential are of the same order. A numerical study cannot answer this, but the recent article \cite{SLQL} explains this.\footnote{In view of the verification of the Luttinger count for electron star spectra in \cite{StelSpec,SLQL}, this had to be so.} 
\item Consistent with the boundary value analysis, there is no evidence of an oscillatory region.
\end{itemize}
The most relevant property of the spectral functions for our question is that as $\kappa$ is increased 
the peak location $k_F$ decreases
orderly and peaks {\em disappear} at various threshold values of $k$. This is the support for our argument that changing $\kappa$ changes the number of microscopic constituents in the electron star. Comparing the the behavior of the various Fermi momenta $k_F$ in the electron star with the results in the extremal AdS-RN black-hole, they are qualitatively identical when one equates $\kap^{-1/2} \sim \qeff$ with the charge of the probe fermion.  
We may therefore infer from our detailed understanding of the behavior of $k_F$ for AdS-RN that also for the electron star as $k_F$ is lowered peaks truly disappear from the spectrum until by extrapolation ultimately one remains: this is the AdS Dirac hair solution \cite{Cubrovic:2010bf}.

%%%%%%%%%%%%%%%%%%%%%%%%%%%%%%%%%%%%%%
%%%%%%%%%%%%%%%%%%%%%%%%%%%%%%%%%%%%%%

We can only make this inference qualitatively as the rapid decrease in spectral weight of each successive peak prevents an exact counting of Fermi surfaces in the numerical results for the electron star spectral functions.  One aspect that we can already see is that as $\kappa$ decreases all present peaks shift to higher $k$, while new
peaks emerging from the left for smaller kappa. This suggests a fermionic version of the UV/IR correspondence where the peak with {\em lowest} $k_F$ corresponds to the last occupied level, i.e. highest ``energy'' in the AdS electron star. We will now address both of these points in more detail.

%%%%%%%%%%%%%%%%%%%%%%%%%%%%%%%%%%%%%%
%%%%%%%%%%%%%%%%%%%%%%%%%%%%%%%%%%%%%%
\section{Fermi surface ordering:  $k_F$  from a Schr\"{o}dinger formulation}
\label{sec:fermi-surf-order}
%%%%%%%%%%%%%%%%%%%%%%%%%%%%%%%%%%%%%%
%%%%%%%%%%%%%%%%%%%%%%%%%%%%%%%%%%%%%%

Our analysis of the behavior of boundary spectral functions as a function
of $\kappa$ relies on the numerically quite evident peaks. Stricly speaking, however, we have not shown that there is a
true singularity in the Green's function at $\ome=0, k=k_F$. We
will do so by showing that the AdS Dirac equation, when recast as
a Schr\"{o}dinger problem has quasi-normalizable solutions at
$\ome=0$ for various $k$. As is well known, in AdS/CFT each such
solution corresponds to a true pole in the boundary Green's
function. Using a WKB approximation for this Schr\"odinger problem
we will in addition be able to estimate the number of poles for a
fixed $\kappa$ and thereby provide a quantitative value for the deviation from the adiabatic background.

We wish to emphasize that the analysis here is
general and captures the behavior of spectral functions in all spherically symmetric and static backgrounds backgrounds alike, whether  AdS-RN, Dirac hair or electron star.

% To figure out the Fermi momentum we will rewrite the Dirac
% equation in the second order form. We need to find the solution
% $y_+$ or $z_+$ which satisfies ingoing boundary condition near the
% horizon and asymptotic to zero near boundary.

%$\bf{z_+~case}:$
\def\hval{\hat{\mu}_{\qeff}}
The $\ome=0$ Dirac equation (\ref{Dirac}) for one set of components (\ref{eq:49}, \ref{eq: dirac2com}) with the replacement $iy_-\to y_-,$ equals
\bea
\sqrt{g_{ii}g^{rr}}\partial_r y_-+\meff\sqrt{g_{ii}}y_-&=&-(k-\hval)z_+,\nonumber\\
\sqrt{g_{ii}g^{rr}}\partial_r z_+-\meff\sqrt{g_{ii}}z_+&=&-(k+\hval)y_-,
\eea where $\hval=\sqrt{\frac{g_{ii}}{-g_{tt}}} \qeff A_t$ and we will drop the subscript $x$ on $k_x$.
%We will focus on $w=0$ in the following. Although this case has the same
% underlying symmmetry as the general one, it turns out to be
% computationally more tractable.
In our conventions $z_+$ (and $y_+$) is the fundamental component dual to the source of the fermionic operator in the CFT \cite{Liu:2009dm,Cubrovic:2009ye}.
Rewriting the coupled first order Dirac equations as a single second order  equation for $z_+$:
\bea\label{eomz+}
&&\partial_r^2z_++{\mathcal{P}}\partial_rz_++{\mathcal{Q}}z_+=0, \non
&&{\mathcal{P}}=\frac{\partial_r(g_{ii}g^{rr})}{2g_{ii}g^{rr}}-\frac{\partial_r \hval}{k+\hval},\non
&&{\mathcal{Q}}=-\frac{\meff\partial_r\sqrt{g_{ii}}}{\sqrt{g_{ii}g^{rr}}}+\frac{\meff\sqrt{g_{rr}}\partial_r \hval}{k+\hval}-\meff^2g_{rr}-\frac{k^2-\hval^2}{g_{ii}g^{rr}}.
\eea
%the first thing one notes is that both $\mathcal{P}$ and
%$\mathcal{Q}$ diverge at some $r=r_\star$ where $\hval\pm k=0$.
%Since $\hval$ is (chosen to be) a positive
%semidefinite function which increases from $\hval=0$ at the horizon, this implies that for
%negative $k<\hval|_{\infty}$ the wavefunction undergoes a qualitative change
%at some point $r=r_\star$. From the Dirac equation, we see that this is the location where $\pa_r z_+(r_{\star}) = \meff \sqrt{g_{rr}}z_+(r_{\star})$.
%Similarly for positive $k$, there is a special location $r_{\star}$ where
%$\pa_r y_-(r_{\star}) = -\meff \sqrt{g_{rr}}z_-(r_{\star})$.
the first thing one notes is that both $\mathcal{P}$ and
$\mathcal{Q}$ diverge at some $r=r_*$ where $\hval + k=0$.
Since $\hval$ is (chosen to be) a positive
semidefinite function which increases from $\hval=0$ at the horizon, this implies that for
negative k (with $-k<\hval|_{\infty}$) the wavefunction is qualitatively different from the wavefunction with positive $k$ which experiences no singularity.%  a qualitative change
% at some point $r=r_*$. From the Dirac equation, we see that (\ref{eomz+}) is not valid at this point and we should have
% $\pa_r z_+(r_{*}) = \meff \sqrt{g_{rr}}z_+(r_{*})$.
% Similarly if one consider $y_+$, for positive $k$, there is also a special location $r_*$ at which
% $\hval - k=0$, and we have the first order equation $\pa_r y_+(r_*) = \meff \sqrt{g_{rr}}y_+(r_*)$.
% Solutions which satisfy both incoming horizon boundary and normalizable boundary conditions at spatial infinity will only occur for specific values of $k$ and through this feature the solution also  singles out a specific scale $r_*$.
% This might be the gravitational perspective on the emergence of a scale in the spectral function. One can avoid this problem by considering a special WKB limit with large $k$. Here for simplicity we will only consider the sector with positive $k$ for  (\ref{eomz+}).
%We will show in a moment how this singularity affects the Dirac equation
The analysis is straightforward if we transform the first derivative away and recast it in the form of a
Schr\"{o}dinger equation by redefining the radial coordinate:
\begin{equation}
\label{tortica}\frac{ds}{dr}=\exp\left(-\int^r dr'
{\mathcal{P}}\right) ~~~\Rightarrow ~~~ s=c_0\int_{r_\infty}^{r}dr'
\frac{|k+\hval|}{\sqrt{g_{ii}g^{rr}}}
\end{equation}
where $c_0$ is an integration constant whose natural scale is of order $c_0 \sim \qeff^{-1}$. This is a simpler version of the generalized $k$-dependent tortoise coordinate
introduced in \cite{Faulkner:2009wj}. In the new
coordinates the equation (\ref{eomz+}) is of the standard form:
\bea
\label{eq:44}
&&\partial_s^2z_+-V(s)z_+=0
\eea
with potential
\bea
&&\label{potz+}
V(s)=-\frac{g_{ii}g^{rr}}{c_0^2|k+\hval|^2}{\mathcal{Q}}.
\eea
\def\muren{m_{\mathtt{ren}}}
%Choosing the integer constant to equal $c_0=1/k$, 
The above potential
(\ref{potz+}) can also be written as
%\begin{equation}
%\label{pot+3}
%V(s)=\frac{1}{(1+\hval/k)^2}\bigg[(k^2+\meff^2 g_{ii}-\hval^2)
%+\meff g_{ii}\sqrt{g^{rr}}\partial_r\ln\frac{\sqrt{g_{ii}}}{k+\hval}\bigg].
%\end{equation}
\begin{equation}
\label{pot+3}
V(s)=\frac{1}{c_0^2(k+\hval)^2}\bigg[(k^2+\meff^2 g_{ii}-\hval^2)
+\meff g_{ii}\sqrt{g^{rr}}\partial_r\ln\frac{\sqrt{g_{ii}}}{k+\hval}\bigg].
\end{equation}

We note again the potential singularity for negative $k$, but before we discuss this
we first need the boundary conditions. The universal boundary behavior is at spatial infinity and follows from the asymptotic AdS geometry. In the adapted coordinates $r\to \infty$ corresponds to $s\to 0$ as follows from
$ds/dr\simeq c_0(k+\hval|_{\infty})/r^2$. The potential therefore equals
\begin{equation}
\label{bndtor}V(s)\simeq \frac{1}{s^2}\big(\meff+\meff^2\big)+\ldots
\end{equation}
and the asymptotic behavior of the two independent solutions equals $z_+ =a_1 s^{-\meff}+b_1 s^{1+\meff}+\ldots$. The second solution is normalizable and we thus demand $a_1=0$.

In the interior, the near-horizon geometry generically is Lifshitz
\begin{eqnarray}
  \label{eq:22}
  ds^2=-r^{2z}dt^2+\frac{1}{r^2}dr^2+r^2(dx^2+dy^2) +\ldots,~~~A=h_{\infty}r^zdt+\ldots,
\end{eqnarray}
with finite dynamical critical exponent $z$ --- AdS-RN, which can be viewed as a special case where
$z\to\infty$, will be given separately. In adapted coordinates
% $r\to 0$, $ds/dr\simeq c_0k\sqrt{g_\infty}/r^2$, and
the interior $r\to 0$ corresponds to  $s\to
-\infty$ and it is easy to show that in this limit potential behaves as
\begin{equation}
\label{hortor}
V(s)\simeq
\frac{1}{c_0^2}+\frac{1}{s^2}\big(\meff\sqrt{g_\infty}+\meff^2g_\infty-h_{\infty}^2\qeff^2g_{\infty}\big)+\ldots.
\end{equation}
Near the horizon the two independent solutions for the wavefunction $z_+$ therefore behave as
\bea
&&z_+\to a_0 e^{-s/c_0}+b_0 e^{s/c_0}.
\eea
The decaying solution $a_0=0$ is the normalizable solution we seek.

\begin{figure}
(A)
\includegraphics[width=0.4\textwidth]{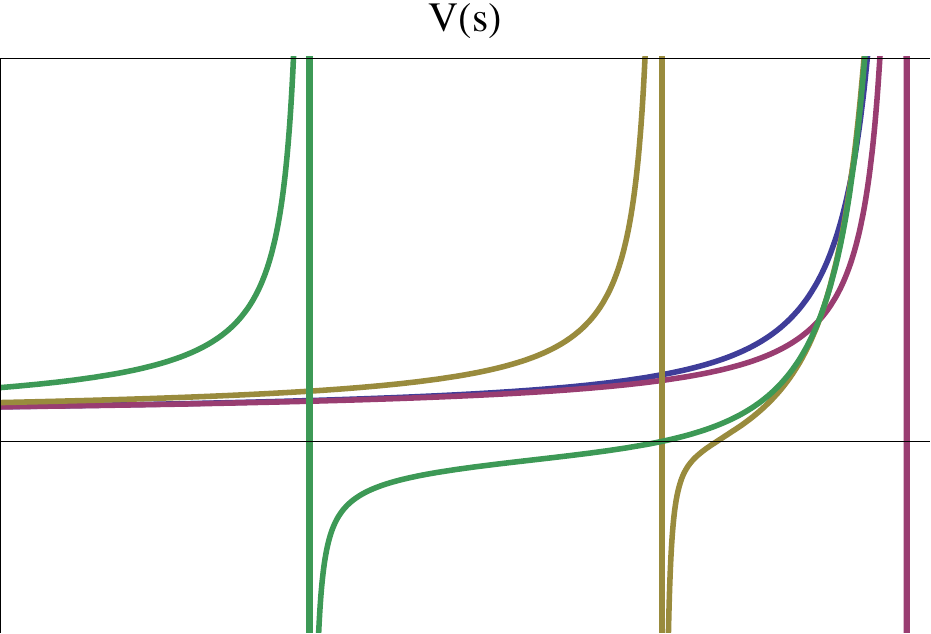}
(B)
\includegraphics[width=0.4\textwidth]{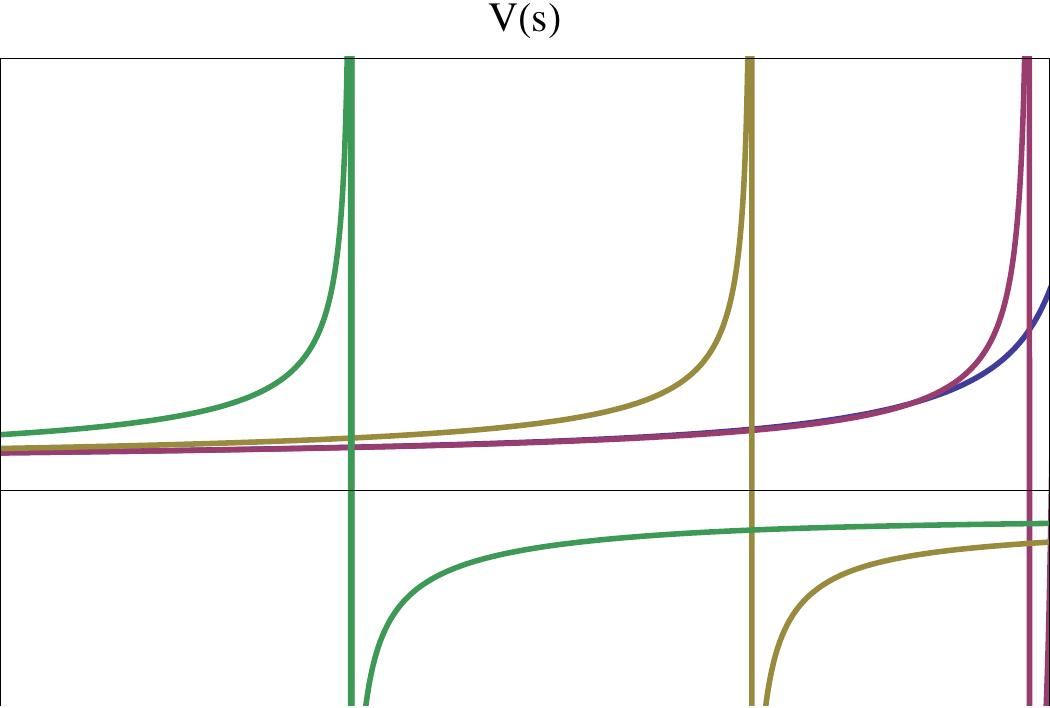}
\caption{
\label{potNegk}
The behavior of the Schr\"{o}dinger potential $V(s)$ for $z_+$ when $k$ is negative. Such a potential has no zero-energy bound state. The potential is rescaled to fit on a finite range. As $|k|$ is lowered below $k_{max}$ for which the potential is strictly positive, a triple pole appears which moves towards the horizon on the left (Fig A. The Blue,Red,Orange,Green curves are decreasing in $|k|$). The  pole hits the horizon for $k=0$ and disappears. Fig B. shows the special case $\meff=0$ where two zeroes collide with two of the triple poles to form a single pole.
}
\end{figure}

Let us now address the possible singular behavior for $k<0$.
To understand what happens, let us first analyze the potential qualitatively for positive $k$. Since the potential is positive semi-definite at the horizon and the boundary, the Schr\"odinger system (\ref{eq:44}) only has a zero-energy normalizable solution if $V(s)$ has a range $s_1<s<s_2$ where it is negative.  This can only at locations where $k^2 < \hval^2-\meff^2g_{ii}-\meff g_{ii}\sqrt{g^{rr}}\partial_r\ln\frac{\sqrt{g_{ii}}}{k+\hval}$.  Defining a ``renormalized'' position dependent mass $\muren^2=\meff^2g_{ii}+\meff g_{ii} \sqrt{g^{rr}}\partial_r\ln\frac{\sqrt{g_{ii}}}{k+\hval}$ this is the intuitive statement that the momenta must be smaller than the local chemical potential $ k^2 < \hval^2-\muren^2$. For positive $k$ the saturation of this bound $k^2 = \hval^2-\muren^2$ has at most two solutions, which are regular zeroes of the potential. This follows from the fact that $\hval^2$ decreases from the boundary towards the interior. If the magnitude $|k|$ is too large the inequality cannot be satisfied, the potential is strictly positive and no solution exists.
For negative $k$, however, the potential has in addition a triple pole at $k^2 =\hval^2$; two poles arise from the prefactor and the third from the $\meff\pa_r\ln(k+\hval)$ term. This pole always occurs closer to the horizon than the zeroes and the potential therefore qualitatively looks like that in Fig. \ref{potNegk} (Since $\hval$ decreases as we move inward from the boundary, starting with $\hval^2>\hval^2-\mu^2 >k^2$, one first saturates the inequality that gives the zero in the potential as one moves inward.) Such a potential cannot support a zero-energy bound state, i.e. eq. (\ref{eq:44}) has no solution for negative $k$. In the case $\meff=0$ a double zero changes the triple pole to a single pole and the argument still holds. This does not mean that there are no $k<0$ poles in the CFT spectral function. They arise from the other physical polarization $y_+$ of the bulk fermion $\Psi$. From the second set of decoupled first order equations for the other components of the Dirac equation (after replacing $iz_-\to z_-,$)
\bea
\sqrt{g_{ii}g^{rr}}\partial_r y_+-\meff\sqrt{g_{ii}}y_+&=&-(k-\hval)z_-,\nonumber\\
\sqrt{g_{ii}g^{rr}}\partial_r z_-+\meff\sqrt{g_{ii}}z_-&=&-(k+\hval)y_+,
\eea
and the associated second order differential EOM for $y_+$:
\bea\label{eomy+}
&&\pa^2_ry_++{\mathcal{P}}\pa_r y_++{\mathcal{Q}}=0, \non
&&{\mathcal{P}}=\frac{\partial_r(g_{ii}g^{rr})}{2g_{ii}g^{rr}}-\frac{\partial_r \hval}{-k+\hval},\non
&&{\mathcal{Q}}=-\frac{\meff\partial_r\sqrt{g_{ii}}}{\sqrt{g_{ii}g^{rr}}}+\frac{\meff\sqrt{g_{rr}}\partial_r \hval}{-k+\hval}-\meff^2g_{rr}-\frac{k^2-\hval^2}{g_{ii}g^{rr}},
\eea
one sees that the Schr\"odinger equation for $y_+$ is the $k\to -k$ image of the equation (\ref{eq:44}) for $z_+$ and thus $y_+$ will only have zero-energy solutions for $k<0$. For simplicity we will only analyze the $z_+$ case.
Note that this semi-positive definite momentum structure of the poles is a feature of any AdS-to-Lifshitz metric different from AdS-RN, where one can have negative $k$ solutions \cite{Faulkner:2009wj}.

The exact solution of (\ref{eq:44}) with the above boundary
conditions corresponding to poles in the CFT spectral function is difficult to find. By construction the system is however equivalent to a Schr\"{o}dinger problem of finding a zero energy solution $z_+$ in the potential (\ref{potz+}) and can be solved in the WKB approximation (see e.g. \cite{Faulkner:2009wj,Roberts}). The WKB approximation holds when $|\partial_s V|\ll |V|^{3/2}$. Notice that this is more general than the background adiabacity limit $\meff \gg 1, \qeff \gg 1$ with $\hat{\beta},~\hat{m}$ fixed % of the background
%.  The WKB approximation applied to {solving the
% Schr\"{o}dinger equation for the probe} is not directly equivalent with an adiabaticity condition on
% the background,
. 
Combining background adiabaticity with a scaling limit $k \gg 1, \meff \gg 1, \qeff \gg 1$ with $c_0k$ fixed and $k$ is parametrically larger than $\hval$ one recovers the WKB potential solved in \cite{StelSpec,SLQL}. As our aim is to study the the deviation away from the background adiabatic limit we will be more general and study the WKB limit of the potential itself, without direct constraints on $\qeff, \meff$. And rather than testing the inequality $|\pa_s V| \ll |V|^{3/2}$ directly, we will rely on the rule of thumb that the WKB limit is justified
when the number of nodes in the wave-function is large. We will
therefore estimate the number $n$ of bound states and use $n\gg 1$
as an empirical justification of our approach.\footnote{A large number of bound states $n$ implies $|\pa_sV|\ll |V|^{3/2}$ if the potential has a single minimum, but as is well known there are systems, e.g. the harmonic oscillator, where the WKB approximation holds for small $n$ as well.}
With this criterion we will be able to study the normalizable solutions to the Dirac equation/pole structure of the CFT spectral functions as a function of $\kappa/L$. %  Notice however that
% in some cases we will find WKB to be a well-behaving limit also
% when $n$ is of order unity.

The potential is bounded both in the AdS boundary and at the horizon, and decreases towards  intermediate values of $r$. We therefore have a standard WKB solution consisting of three regions:
\begin{itemize}
\item
 In the %outer
 regions where $V>0$, the
solution decays exponentially: \be z_+=c_{1,2}V^{-1/4}
{\mathrm{exp}}\bigg(\pm\int_{r_{1,2}}^{r}dr'\big[c_0\sqrt{g^{ii}g_{rr}}\big(k+\hval\big)\sqrt{V}\big]\bigg).\ee
Here $r_1,~r_2$ are the turning points where $V(r_1)=0=V(r_2)$.
% To match the boundary condition (\ref{bndcod}), we need to pick
% the $``-"$ sector for $r_2$.

\item In the region $r_1<r<r_2$, i.e. $V<0$, the solution is
\be
z_+=c_{3}(-V)^{-1/4}{\mathrm{Re}}
\bigg[{\mathrm{exp}}\big(^{i\int_{r_{1}}^{r}dr'[c_0\sqrt{g^{ii}g_{rr}}\big(k+\hval\big)\sqrt{-V}]-i\pi/4}\big)\bigg],
\ee
with the constant phase $-i\pi/4$ originating in the connection formula at
the turning point $r_1$.
\end{itemize}

Finding a WKB solution shows us that the peaks seen numerically are true poles in the spectral function. But it also allows us to estimate the number of peaks that the numerical approach could not resolve.
The WKB quantization condition
\be
\int_{r_{1}}^{r_2}dr'\bigg[c_0\sqrt{g^{ii}g_{rr}}\big(k+\hval\big)\sqrt{-V}\bigg]=\pi(n+1/2)
\ee
counts the number of bound states with negative semi-definite energy.
Note that $n$ does not depend on the integral constant as there is also a factor $1/c_0$ in $\sqrt{-V}$.
 Since $V$ depends on $k$, we will see that as we increase $k$ this number decreases. The natural interpretation in the context of a bulk many-body Fermi system is that this establishes the ordering of the the filling of all the $\ome=0$ momentum shells in the electron star. For a fixed $k$ one counts the modes that have been previously occupied and, consistent with our earlier deduction,  the lowest/highest $k_F$ corresponds to the last/first occupied state. Though counterintuitive from a field theory perspective where normally $E \sim k_F$, this UV/IR correspondence is very natural from the AdS-bulk, if one thinks of the electron star as a trapped electron gas. The last occupied state should then be the outermost state from the center, but this state has the lowest effective chemical potential and hence lowest $k_F$.

Let us now show this explicitly by analyzing the potential and the bound states in the electron star and AdS-RN.

\subsubsection*{Electron star}

\begin{figure}[t!]
\begin{center}
\begin{tabular}{cc}
(A)
\includegraphics[width=0.45\textwidth]{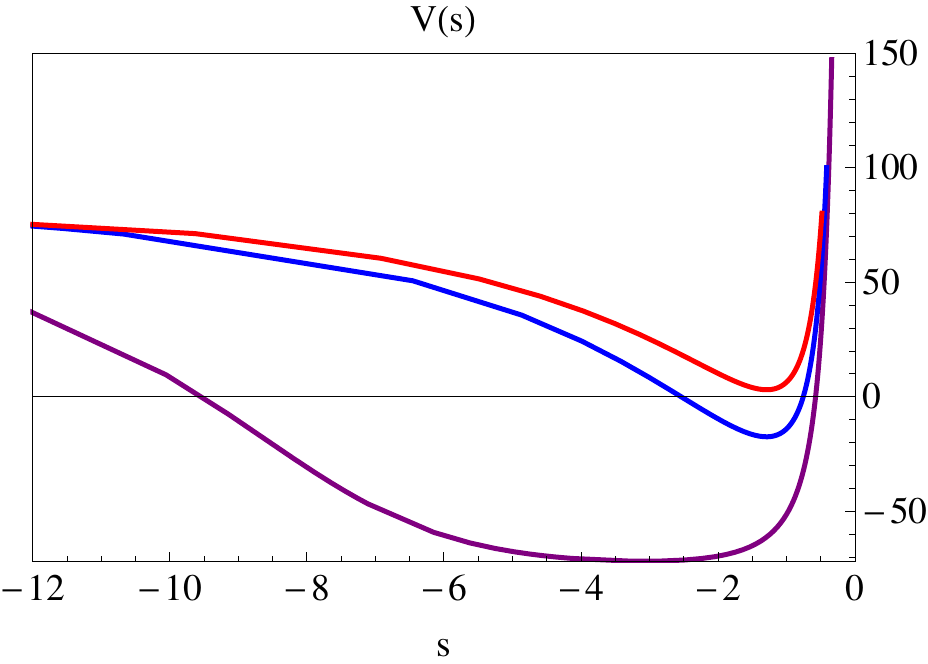}
(B)
\includegraphics[width=0.45\textwidth]{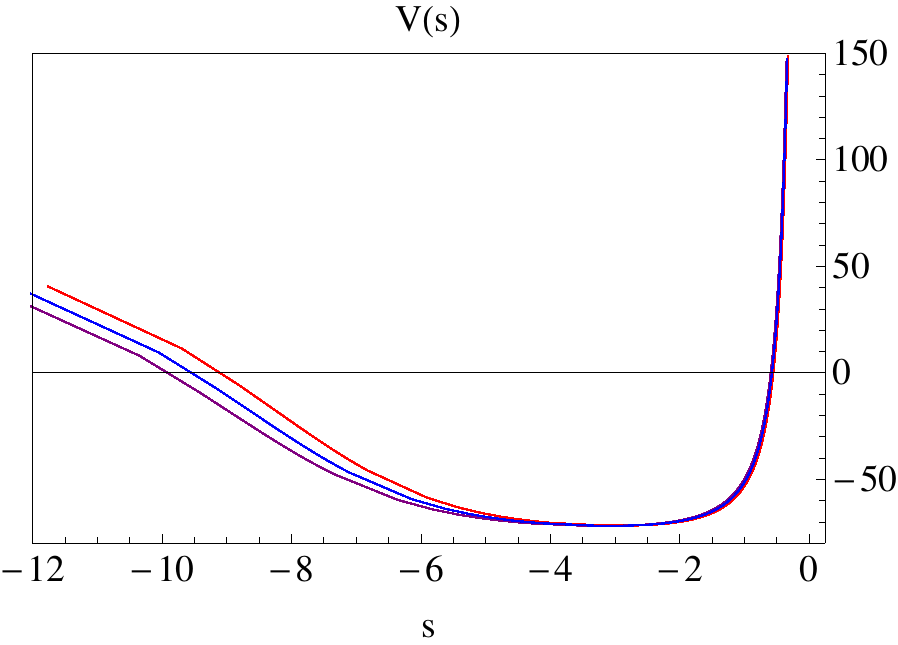}
\end{tabular}
\caption{\small The Schr\"{o}dinger potential $V(s)$ for the fermion component
$z_+$ of in the ES background $\hat{m}=0.36, z=2, c_0=0.1$. Fig. A. shows the dependence on the momentum $k= 0.0185$ ({Purple}), $k = 5$ ({Blue}), $k = 10$ ({Red}) for $\kappa=0.092$. Fig. B. shows the dependence on  $\kappa=0.086$ ({Purple}), $\kappa=0.092$
({Blue}), $\kappa=0.1$ ({Red}) for $k=0.0185$. Recall that $s=0$ is the AdS boundary and $s=-\infty$ is the near-horizon region.} \label{pot}
\end{center}
\end{figure}

\begin{figure}[t!]
\begin{center}
\begin{tabular}{cc}
(A)
\includegraphics[width=0.28\textwidth]{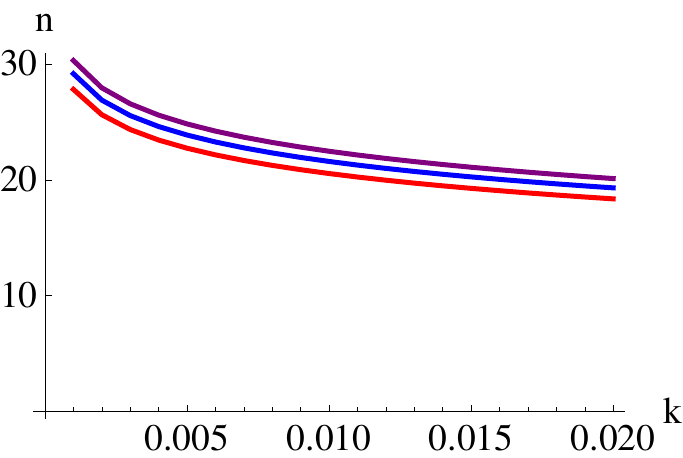}
(B)
\includegraphics[width=0.28\textwidth]{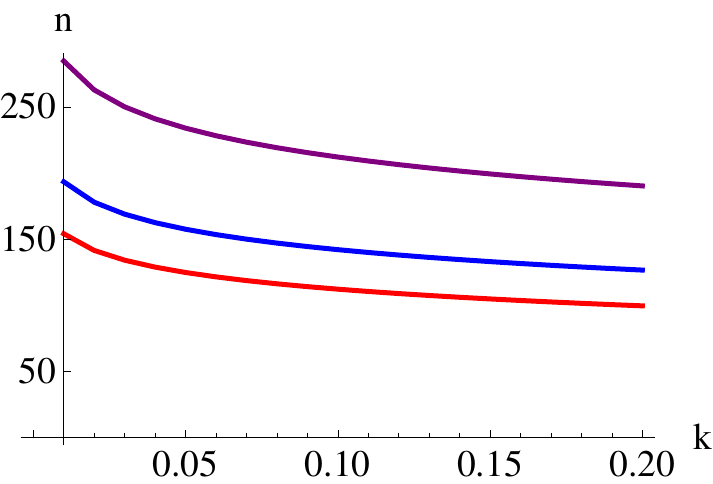}
(C)
\includegraphics[width=0.28\textwidth]{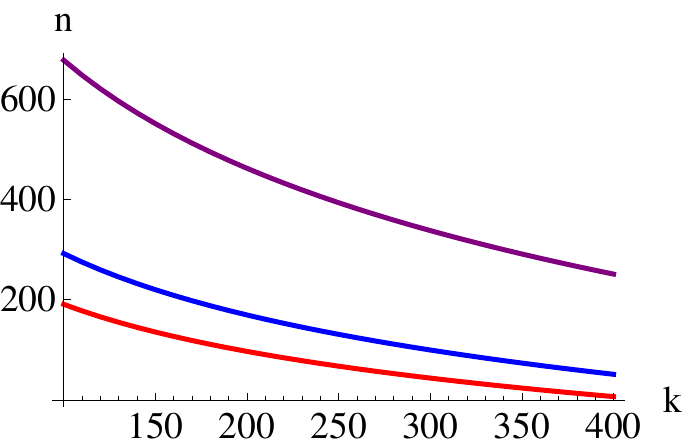}
\end{tabular}
\caption{\small The WKB estimate of the number of bound states $n$ as a function of the momentum $k$ for  $\kappa=0.086
(\mathrm{Purple}),~0.092 (\mathrm{Blue}),~0.1(\mathrm{Red})$ (Fig A.);  for $\kappa=0.001
(\mathrm{Purple}),~0.002 (\mathrm{Blue}),~0.003(\mathrm{Red})$ (Fig B.) and  for $\kappa=10^{-5}
(\mathrm{Purple}),~3\times10^{-5} (\mathrm{Blue}),~5\times10^{-5}
(\mathrm{Red})$ (Fig C.). Note the parametric increase in number of states as the adiabaticity of the background improves for smaller $\kappa$. Both figures are for the electron star background with $\hat{m}=0.36, z=2$. Since $n\gg 1$ in all cases, WKB gives a valid estimate.
} \label{nk1}
\end{center}
\end{figure}

The potential \eqref{pot+3} for the electron star is given in Fig.~\ref{pot} and the number of bound states as a function of $k$ in Fig.~\ref{nk1}. As stated the number of states decreases with increasing $k$, consistent with the analogy of the pole distribution of the spectral functions compared with AdS-RN. Moreover, we clearly see the significant increase in the number of states as we decrease $\kappa/L$ thereby improving the adiabaticity of the background. This vividly illustrates that the adiabatic limit corresponds to a large number of constituents. As all numbers of states are far larger than one, the use of the WKB is justified.

%%%%%%%%%%%%%%%%%%%%%%%%%%%%%%%%%%%%%%
%%%%%%%%%%%%%%%%%%%%%%%%%%%%%%%%%%%%%%

%%%%%%%%%%%%%%%%%%%%%%%%%%%%%%%%%%%%%%
%%%%%%%%%%%%%%%%%%%%%%%%%%%%%%%%%%%%%%

\subsubsection*{The Reissner-Nordstr\"{o}m case}

For AdS-RN the Schr\"odinger analysis requires a separate
discussion of the near horizon boundary conditions, which we present here for completeness and comparison. Part of this analysis is originally worked out in \cite{Faulkner:2009wj}. The
AdS-RN black hole with metric \bea
ds^2&=&L^2\left(-f(r)dt^2+\frac{dr^2}{f(r)}+r^2(dx^2+dy^2)\right),\\
f(r)&=&r^2\bigg(1+\frac{3}{r^4}-\frac{4}{r^3}\bigg),\\
A&=&\mu\bigg(1-\frac{1}{r}\bigg)dt,
\eea
has near horizon geometry AdS$_2 \times \RR^2$
\bea
ds^2&=&-6(r-1)^2dt^2+\frac{dr^2}{6(r-1)^2}+(dx^2+dy^2),\\
A&=&\sqrt{3}\bigg(r-1\bigg)dt.
\eea
A coordinate redefinition of $r$ in eq. \eqref{eq:22} to $r =(r_{AdS_2}-1)^{1/z}$ shows that this  corresponds to a dynamical critical exponent $z=\infty$ and is outside the validity of the previous analysis.

Before we proceed, recall that the existence of AdS$_2\times \RR^2$ near-horizon region allows for a semi-analytic determination of the fermion spectral functions with the self-energy $\Sig \sim \ome^{2\nu_{k_F}}$  controlled by the IR conformal dimension
$\delta_k=1/2+\nu_k$ with
\be \nu_k=\frac{1}{\sqrt{6}}\sqrt{m^2+k^2-\frac{q^2}{2}}~.
\ee
When $\nu_k$ is imaginary, which for $q^2>2m^2$ always happens for small $k$, the spectral function exhibits oscillatory behavior, but generically has finite weight at $\ome=0$. When $\nu_k$ is real,
there are poles in the spectral functions at a finite number of different Fermi momenta $k_F$. The associated quasiparticles can characterize a non-FL  ($\nu_{k_F}<1/2$), a marginal FL ($\nu_{k_F}=1/2$) or irregular FL ($\nu_{k_F}>1/2$) with linear dispersion but width $\Gam \neq \ome^2$ \cite{Faulkner:2009wj}.

%%%%%%%%%%%%%%%%%%%%%%%%%%%%%%%%%%%%%%
\begin{figure}[t!]
\begin{center}
\begin{tabular}{cc}
(A)
\includegraphics[width=0.45\textwidth]{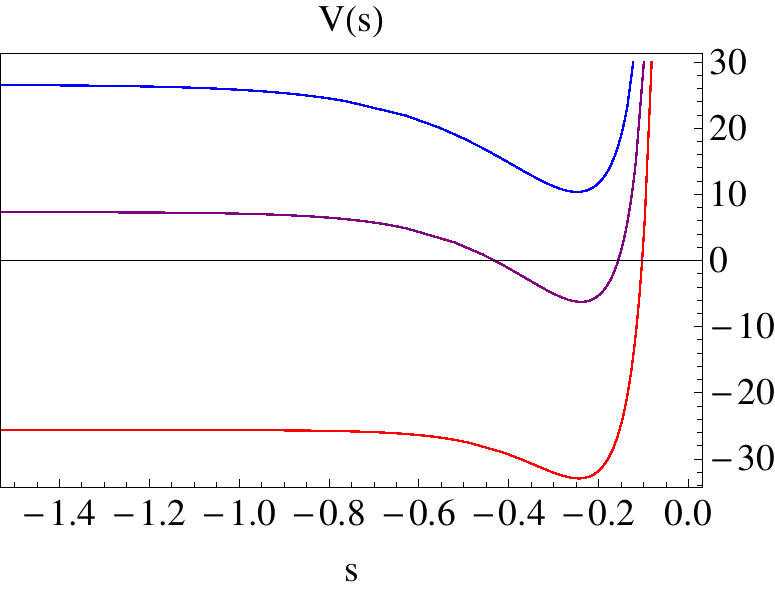}
(B)
\includegraphics[width=0.45\textwidth]{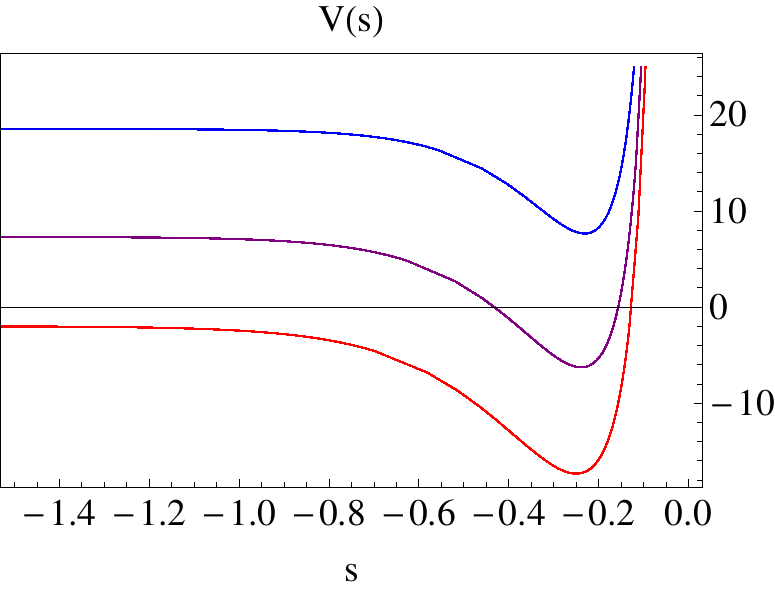}
\end{tabular}
\caption{\small The Schr\"{o}dinger potential $V(s)$ for the fermion component
$z_+$ of in the AdS-RN background $r_+= 1, \mu = \sqrt{3}, g_F = 1, mL = 0.4, c_0=0.1$. Fig. A. shows the dependence on the momentum $k= 1$ ({Red}), $k = 2$ ({Purple}), $k = 3$ ({Blue}) for charge $q=2.5$. Fig. B. shows the dependence on  the charge $q$ --- analogous to $\kappa$ in the ES background ---. Shown are the values $q=2$ ({Blue}), $q=2.5$
({Purple}), $q=3$ ({Red}) for the momentum $k=2$. In both figures the Red potentials correspond to the oscillatory region $\nu_k^2<0$, the Purple potentials show the generic shape that can support an $\ome=0$ bound state, and the Blue potentials are strictly positive and no zero-energy bound state is present. Recall that $s=0$ is the AdS boundary and $s=-\infty$ is the near-horizon region.} \label{potRN}
\end{center}
\end{figure}
%%%%%%%%%%%%%%%%%%%%%%%%%%%%%%%%%%%%%%

The analytic form of the near-horizon metric allows us to solve exactly for the near horizon potential $V$ in terms of
 $s=\frac{c_0}{\sqrt{6}}(k+q/\sqrt{2})\ln{(r-1)}+\ldots$. As noted in \cite{Faulkner:2009wj}
 one remarkably obtains that the near-horizon potential
for $s\to -\infty$ is proportional to the self-energy exponent:
\be
V(s)\simeq \frac{6}{c_0^2(k+q/\sqrt{2})^2}\nu_k^2 +\ldots.
\ee
One immediately recognizes the oscillatory region $\nu_{k}^2<0$ of the spectral function as an $\ome=0$ Schr\"odinger potential which is ``free''  at the horizon $s=-\infty$ (Fig. \ref{potRN}) and no bound state can form. Comparing with our previous results, we see that this oscillatory region is a distinct property of AdS-RN. For any Lifshitz near-horizon metric the potential is always positive-definite near the horizon and all $\ome=0$ solutions will be bounded. (see also \cite{StelSpec,SLQL}). As we increase $k$, $\nu_k^2$ becomes positive, then the AdS-RN potential is also positive at the horizon and bound zero-energy states can form. Increasing $k$ further, one reaches a maximal $k_{max}$, above which the potential is always positive and no zero-energy bound state exists anymore.

\begin{figure}[t!]
\centering
\includegraphics[width=0.5\textwidth]{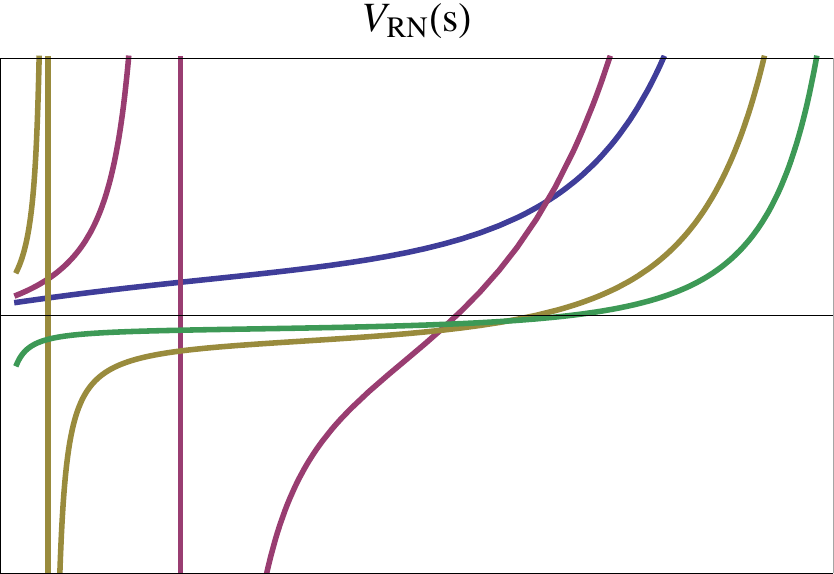}\label{NegKPotRN}
\caption{The qualitative behavior for negative $k$ of the Schr\"{o}dinger potential $V(s)$ for the fermion component
$z_+$ of in the AdS-RN background $r_+= 1, \mu = \sqrt{3}, g_F = 1, mL = 0.1$.
The radial coordinate has been rescaled to a finite domain such that the full potential can be represented in the figure; on the right is the AdS boundary and left is the near-horizon region and the range is slightly extended beyond the true horizon, which is exactly at the short vertical line-segments on the right. Potentials are given for $q=12/\sqrt{3}$, $k=-15$ (Blue) for which the potential is strictly positive, $k=-10$ (Red), $k=-7$ (Orange), which both have triple poles and the pole can be seen to move towards the horizon on the left as $k$ decreases, and $k=-4$ (Green) which has no pole and a finite negative value at the horizon. The pole disappears for $|k|<q/\sqrt{2}$ leaving a regular bounded potential  which can support zero-energy bound states. }
\end{figure}

%%%%%%%%%%%%%%%%%%%%%%%%%%%%%%%%%%%%%%

\begin{figure}[h!]
\begin{center}
\begin{tabular}{cc}
(A)
\includegraphics[width=0.4\textwidth]{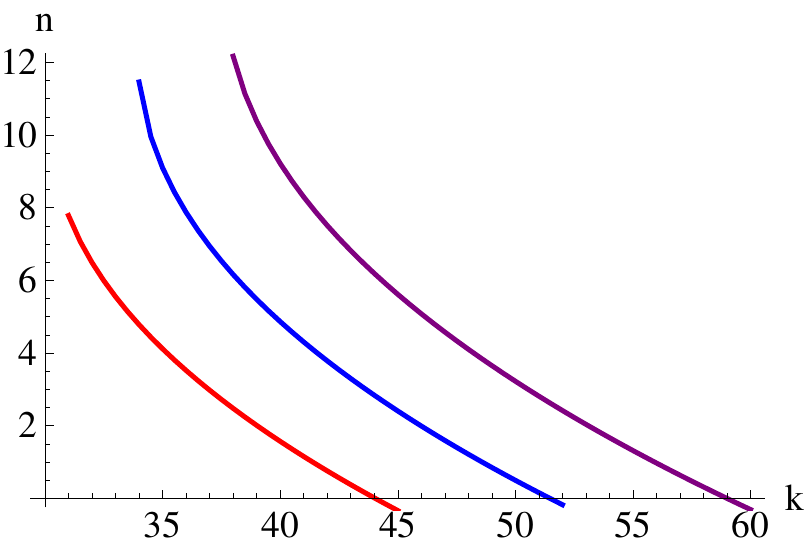}
(B)
\includegraphics[width=0.4\textwidth]{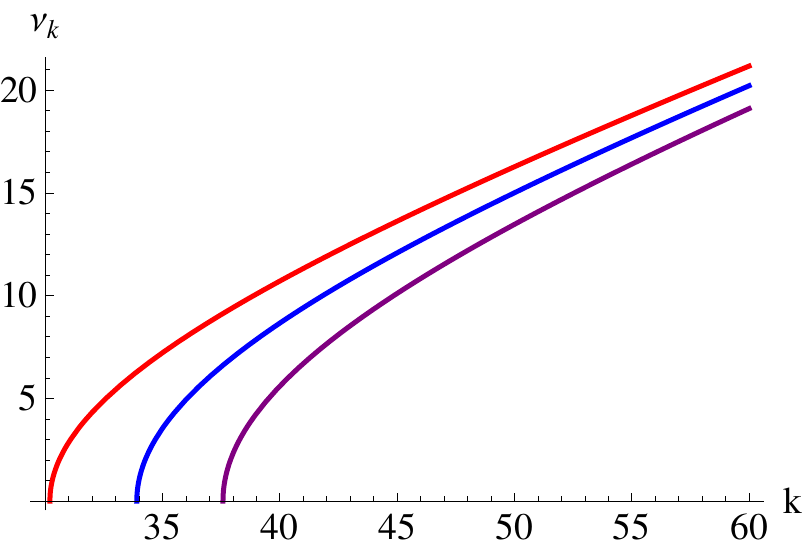}
% (C)
% \includegraphics[width=0.28\textwidth]{wkbncount-highkappa}
%\includegraphics[width=0.5\textwidth]{graph/wkbcon1-eps-converted-to}
\end{tabular}
\caption{\small The WKB estimate of the number of bound states $n$ in the AdS-RN Schr\"odinger potential for $z_+$ with $mL=10$. The WKB approximation only applies to large values of the charge $q = 45$ (Red), $q=50$ (Blue), $q=55$ (Purple). Fig B. gives the associated values of the IR conformal dimension $\nu_k=\frac{1}{\sqrt{6}}\sqrt{m^2+k^2-\frac{q^2}{2}}$.
Both figures are for the extremal AdS-RN background with $\mu=\sqrt{3}, r_+=1, g_F=1$. \label{nk-rn}}
\end{center}
\end{figure}
%%%%%%%%%%%%%%%%%%%%%%%%%%%%%%%%%%%%%%

Because the near-horizon boundary conditions for AdS-RN differ from the general analysis, the possible singularity in the potential for $k<0$ also requires a separate study. This is clearly intimately tied to the existence of an oscillatory regime in the spectral function, as the previous analysis does apply for $\nu_k^2>0$. The clearest way to understand what happens for $\nu_k^2 <0$ is to analyze the potential explicitly. Again if $|k|>k_{max}$ the potential is strictly positive definite, and no zero-energy bound state exists. As we decrease the magnitude of $k<0$, a triple pole will form  near the boundary when $k =-\hval(s)$, soon followed by a zero at $k=-\sqrt{\hval(s)^2-\muren(s)^2}$ (see Fig. \ref{potNegk}). As we approach the horizon, in the general case % $\lim_{s\rar\infty}\hval = h_{\infty}\qeff e^{s/k}+\ldots$ and this pole at $s_{\ast}= k\ln(-k/h_{\infty}\qeff)$  disappears precisely when $k=0$ (for $k<0$ the horizon is at $s=+\infty$). In AdS-RN, however, $\lim_{s\rar \infty} \hval = \frac{q}{\sqrt{2}} e^{s\sqrt{6}/c_0(k+q/\sqrt{2})}+\ldots$, the pole $s_{\ast}^{RN} =\frac{c_0}{\sqrt{6}} (k+\frac{q}{\sqrt{2}})\ln(-k\frac{\sqrt{2}}{q})$
where $\lim_{r\rar 0}\hval = h_{\infty}\qeff r +\ldots$, this pole at $r_{\ast}= -k/(h_{\infty}\qeff)$ hits the horizon and disappears precisely when $k=0$ . In AdS-RN, however, where $\lim_{r\rar 1} \hval = \frac{q}{\sqrt{2}} + \frac{\sqrt{2}q}{3}(r-1)+\ldots$, the pole at $r_{\ast}^{RN}-1 = \frac{3}{\sqrt{2}q}(k+\frac{q}{\sqrt{2}})$ hits the horizon and disappears at $k = -\frac{q}{\sqrt{2}}$. For negative values of $k$ whose magnitude is less than $|k| < \frac{q}{\sqrt{2}}$, the potential is regular and bounded and can and does have zero-energy solutions. Fig. \ref{NegKPotRN} shows this disappearance of the pole for the AdS-RN potential.

Counting solutions through WKB is also more complicated for AdS-RN. For $\cO(1)$ values of $q$ there are only few Fermi surfaces and the WKB approximation does not apply. For large $q$ it does, however. For completeness we show the results in Fig. \ref{nk-rn}.

\section{Conclusion and Discussion}

\begin{figure}[t!]
  \includegraphics[width=0.4\textwidth]{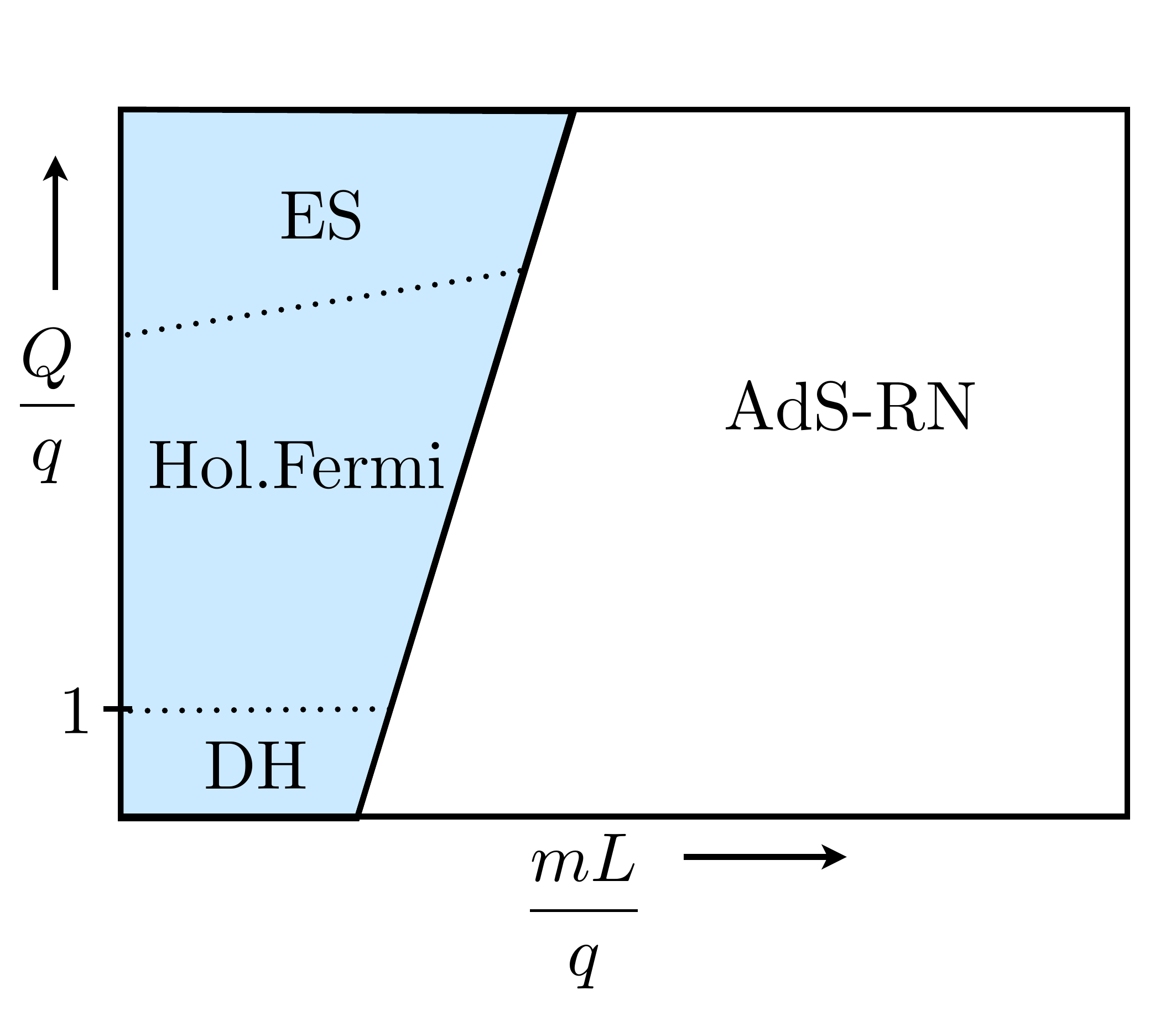}
~
  \includegraphics[width=0.4\textwidth]{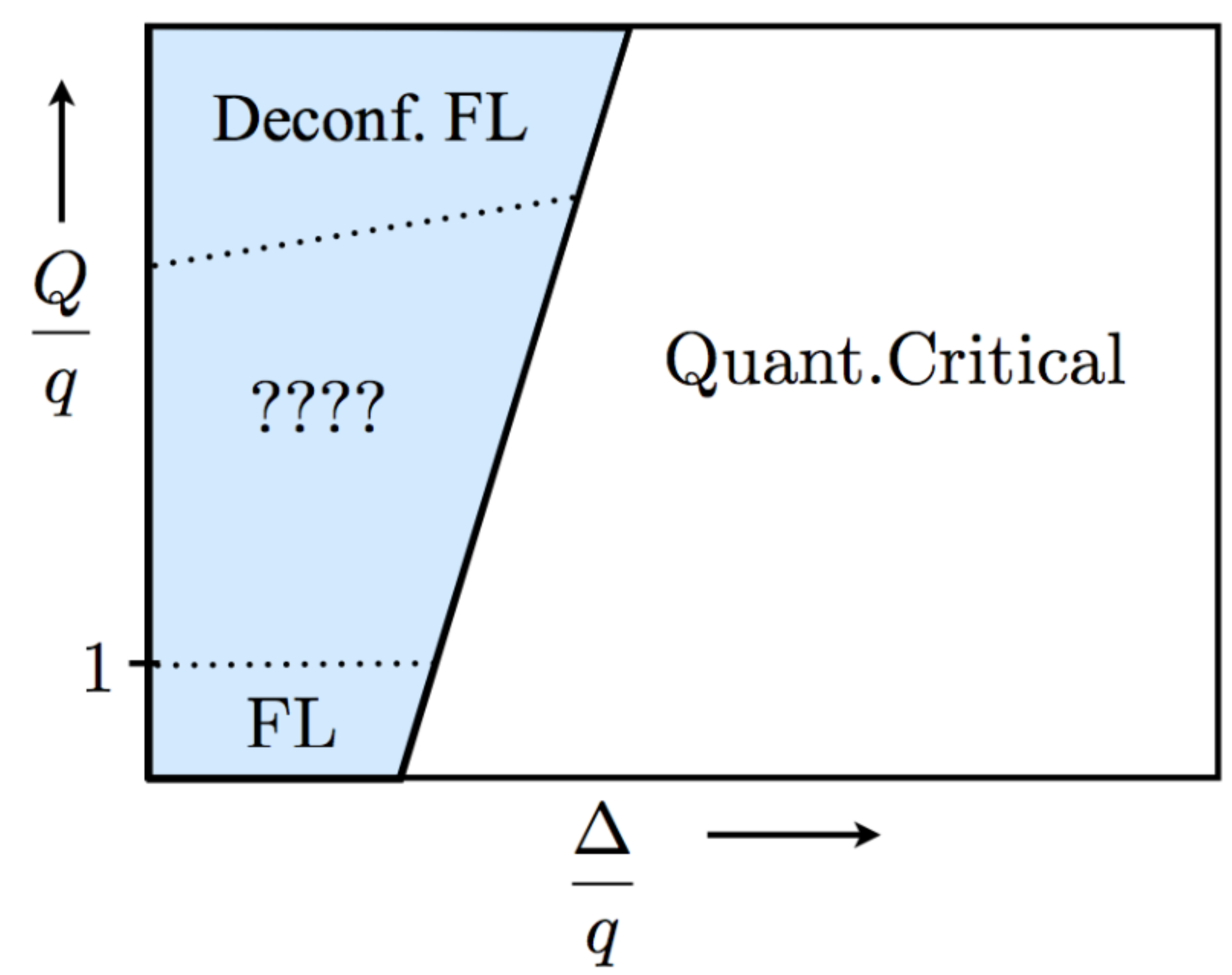}
  \caption{Schematic diagram of the different groundstate solutions of strongly coupled fermions implied by holography for fixed charge density $Q$. Here $q$ is the constituent charge of the fermions and $mL \sim \Delta$ the mass/conformal scaling dimension of the fermionic operator. One has the gravitational electron star (ES)/Dirac Hair (DH) solution for large/small $Q/q$ and small $mL/q$ dual a deconfined Fermi liquid/regular Fermi liquid in the CFT. For $mL/q \sim \Delta/q$ large the groundstate remains the fermionic quantum critical state dual to AdS-RN.}
\label{fig:22}
\end{figure}

These electron star spectral function results answer two of the three questions raised in the introduction directly. 
\begin{itemize}
\item They show explicitly how the fermion wavefunctions in their own gravitating potential well are ordered despite the fact that they all have strictly vanishing energy: In a fermionic version of the UV-IR correspondence they are ordered {\em inversely} in $k$, with the ``lowest''/first occupied state having the highest $k$ and the ``highest''/last occupied state having the lowest $k$. With the qualitative AdS/CFT understanding that scale corresponds to distance away from the interior, one can intuitively picture this as literally filling geometrical shells of the electron star, with the outermost/highest/last shell at large radius corresponding to the wavefunction with lowest local chemical potential and hence lowest $k$.
\item
The decrease of the number of bound states --- the number of occupied wavefunctions in the electron star --- as we decrease $\qeff = \hat{\beta}^{1/4}\sqrt{\frac{\pi L}{\kappa}}$ for a fixed electron star background extrapolates naturally to a limit where the number of bound states is unity. This extrapolation pushes the solution beyond its adiabatic regime of validity. In principle we know what the correct description in this limit is: it is the AdS Dirac Hair solution constructed in \cite{Cubrovic:2010bf}. The dependence of the number of bound states on $\kappa/L$ therefore illustrates that the electron star and Dirac Hair solutions are two limiting cases of the gravitationally backreacted Fermi gas. 
\end{itemize}
With this knowledge we can schematically classify the groundstate solutions of AdS Einstein-Maxwell gravity minimally coupled to charged fermions at finite charge density. For large mass $mL$ in units of the constituent charge $q$, the only solution is a charged AdS-Reissner-N\"ordstrom black hole. For a low enough mass-to-charge ratio, the black hole becomes unstable and develops hair. If in addition the total charge density $Q$ is of the order of the microscopic charge $q$ this hairy solution is the Dirac Hair configuration constructed in \cite{Cubrovic:2010bf}, whereas in the limit of large total charge density $Q$ one can make an adiabatic Thomas Fermi approximation and arrives a la Tolman-Oppenheimer-Volkov at an electron star (Fig. \ref{fig:22}).

Translating this solution space through the AdS/CFT correspondence one reads off that in the dual strongly coupled field theory, one remains in the critical state if the ratio of the scaling dimension to the charge $\Delta/q$ is too large. For a small enough value of this ratio, the critical state is unstable and forms a novel scaleful groundstate. The generic condensed matter expectation of a unique Fermi liquid is realized if the total charge density is of the same order as the constituent charge. 
Following  \cite{StelSpec,SLQL} and \cite{Hartnoll:2010xj,Sachdev:2010um,Huijse:2011hp} the state for $Q \gg q$ is some deconfined Fermi liquid.

The gravity description of either limit has some deficiencies, most notably the lack of an electron star wavefunction at infinity and the unnatural restriction to $Q=q$ for the Dirac Hair solution. A generic solution for $Q\geq q$ with wavefunction tails extending to infinity as the Dirac hair would be a more precise holographic dual to the strongly interacting large $N$ Fermi system. Any CFT information can then be cleanly read off at the AdS boundary. A naive construction could be to superpose Dirac Hair onto the electron star; in principle one can achieve this solution by a next order Hartree-Fock or  Local Density Approximation computation.

This best-of-both-worlds generic solution ought to be the {true holographic dual of the strongly interacting Fermi ground state}. If one is able to answer convincingly how this system circumvents the wisdom that the groundstate of an interacting many-body system of fermions is a generic single quasiparticle Landau Fermi liquid, then one would truly have found {a finite density Fermi system that does not refer at any stage to an underlying perturbative Fermi gas.}

% answer two out of the three questions raised in the introduction directly and give an indication towards the answer of the 
 
% Having argued that the behavior of the spectral functions as a
% function of $\kappa$ shows that the electron star and Dirac Hair
% solution are two limits of the same system, we have in principle
% given a resolution.

% \begin{itemize}
% \item Overarching question: what is the holographic Fermi groundstate.
% Fermi Ball/Frac FL/reg FL
% % \item Ordering in ES: answered (UV-IR)
% % \item AdS Boundary of ES: DH\&ES opposing limits, truth in middle.
% \end{itemize}

\section*{Acknowledgements}
\noi
We thank S. Hartnoll, A. Karch, H. Liu, T. K. Ng and B. Overbosch for discussions and correspondence.
% We are very grateful to the KITP Santa Barbara for
% its generous hospitality and the organizers and participants of
% the AdS/CMT Workshop (July 2009).
KS is very grateful to the Hong Kong Institute for Advanced Studies for the hospitality during the completion of this work.
This research was supported in part by a VIDI Innovative Research
Incentive Grant (K. Schalm) from the Netherlands Organisation for
Scientific Research (NWO), a Spinoza Award (J. Zaanen) from the
Netherlands Organisation for Scientific Research (NWO) and the
Dutch Foundation for Fundamental Research on Matter (FOM).

\end{document}